\documentclass[prl, aps, twocolumn, superscriptaddress,showpacs]{revtex4}
\usepackage{eqnarray}
\usepackage[english]{babel} 
\usepackage[english]{layout}
\usepackage{amsmath,amsfonts,amssymb}
\usepackage{graphicx}
\usepackage{float}
\usepackage{color}
\usepackage{braket}
\usepackage{version}
\usepackage{array,multirow,makecell}

\begin{document}
\title{Edge-emitting polariton laser and amplifier based on a ZnO waveguide}
\author{O. Jamadi}
\author{F. R\'everet}
\author{P. Disseix}
\author{F. M\'edard}
\author{J. Leymarie}
\author{A. Moreau}
\author{D. Solnyshkov}
\affiliation{Institut Pascal, PHOTON-N2, University Clermont Auvergne, CNRS, 4 avenue Blaise Pascal, 63178 Aubi\`{e}re Cedex, France.} 
\author{C. Deparis}
\author{M. Leroux}
\affiliation{UCA, CRHEA-CNRS, Valbonne, France} 
\author{E. Cambril}
\author{S. Bouchoule}
\affiliation{Centre Nanosciences et Nanotechnologies (C2N), CNRS, Université Paris-Sud, Marcoussis, France.} 
\author{J. Zuniga-Perez}
\affiliation{UCA, CRHEA-CNRS, Valbonne, France} 
\author{G. Malpuech}
\affiliation{Institut Pascal, PHOTON-N2, University Clermont Auvergne, CNRS, 4 avenue Blaise Pascal, 63178 Aubi\`{e}re Cedex, France.}

\begin{abstract}
We demonstrate edge-emitting exciton-polariton (polariton) lasing from 5 to 300 K and amplification of non-radiative guided polariton modes within ZnO waveguides. The mode dispersion below and above the lasing threshold is directly measured using gratings present on top of the sample, fully demonstrating the polaritonic nature of the lasing modes. The threshold is found to be similar to that of radiative polaritons in planar ZnO microcavities. These results open broad perspectives for guided polaritonics by allowing an easier and more straightforward implementation of polariton integrated circuits  exploiting fast propagating polaritons.
\end{abstract}
\maketitle

Exciton-polaritons (polaritons) are quasi-particles resulting from the coupling between a light mode and an excitonic resonance. They have been theoretically introduced by Hopfield and Agranovich \cite{Hopfield1958, Agranovich1959} at the end of the 50's to describe light propagation in bulk semiconductors. In 1992, the achievement of strong light-matter coupling between the photonic radiative modes of a planar microcavity and the excitonic resonances of embedded quantum wells \cite{Weisbuch1992} opened the era of 2-dimensional cavity polaritons, which since then have been studied very extensively. From the fundamental side, polaritons represent a  direct implementation of a spinor quantum fluid of light \cite{Shelykh2010,Carusotto2013} with a unique access to the time-dependent wavefunction in real and reciprocal space via optical measurements. It allowed the study of Bose-Einstein Condensation in open systems \cite{Richard2005,Kasprzak2006,Kasprzak2008,Sun2017}, of the superfluidity of light \cite{Amo2009,Sanvitto2016}, of various types of topological defects \cite{Lagoudakis2008,Lagoudakis2009,Hivet}, of the creation of spin flows \cite{Kavokin2005,Leyder2007}. In-plane potentials are commonly realized for a decade. They allow to implement coupled 0D polariton modes, building artificial molecules \cite{Galbiati2012,Sala2014} or lattices \cite{Lai2007,Jacqmin2014,Kim2014,Whittaker2017}. This opens new perspectives for emulating different physical systems such as topological insulators of various types\cite{Nalitov2014b,Bleu2017z} or the classical XY model \cite{Berloff2017}. From the applied side, polaritons have the critical advantage of their high nonlinear response, low threshold operation, and potentially high scalability \cite{Sanvitto2016}, useful for realizing low-consumption, compact, all-optical devices, which could replace electronics for some tasks \cite{Caulfield2010}. 

The most paradigmatic polariton device is the so-called polariton laser \cite{Imamoglu1996},  based on polariton condensation. It does not require electron-hole gain and can therefore exhibit a very low threshold. In GaAs-based samples, at 5K, a wide variety of devices such as switches and optical transistors have been demonstrated\cite{Amo2010,Gao2012,Wertz2012,Ballarini2013,Nguyen2013,Sanvitto2016}. Most of them are based on the creation of a polariton flow based on slow radiative modes propagating at 1-2\% of the speed of light. This is only possible at low temperature in very high Q samples (typically $Q\propto 10^5$). Room-temperature polaritonics requires the use of microcavities based on large-bandgap semiconductors (GaN, ZnO) or various organic materials, where polariton lasing has also been demonstrated  \cite{Christopoulos2007,Christmann2008,Bhattacharya2014,Feng2013,Cohen2010,Dietrich2016}. However, their Q-factors remain limited to a few thousands, which makes difficult the implementation of room-temperature polariton-based switches exploiting the concepts developed in GaAs-based samples.
 
An alternative geometry for polaritonics is the one where either Bloch surface waves \cite{Liscidini2011,Pirotta2014,Lerario2017} or guided modes confined by total internal reflection in a layer \cite{Solnyshkov2014,Walker2013,Rosenberg2016,Ellenboger2011,Ciers2017,Hu2017,Walker2015} strongly couple to excitonic resonances. The geometry is very appealing because of its simple technological realization, the easier electrical injection, and the possibility it opens to design integrated polaritonic circuits with very limited radiative losses. Guided polaritons have been observed and studied in GaAs \cite{Walker2013,Rosenberg2016}, organic materials \cite{Pirotta2014,Ellenboger2011}, GaN \cite{Ciers2017}, and Transitional Metal Dichalcogenides (TMD) \cite{Hu2017}, with recent reports of nonlinear polariton-polariton interaction \cite{Lerario2017,Walker2015}. However, even if theoretically predicted \cite{Solnyshkov2014}, "horizontal" edge-emitting polariton lasing in such fast-propagating polariton modes has never been observed.

In this work, we achieve one of the crucial milestones of polaritonics by reporting lasing from 5 to 300 K and amplification of guided polariton modes under optical pumping in ZnO-based waveguides. The mode dispersion below and above the lasing threshold is directly measured using gratings present on top of the sample. These dispersions clearly demonstrate the polaritonic nature of the lasing modes.

We use two ZnO-ZnMgO waveguides grown by Molecular Beam Epitaxy on m-plane bulk ZnO substrates. The thicknesses of the active ZnO layers are 50 nm (W1 sample) and 130 nm (W2 sample), respectively. A sketch of W1 is shown in Fig.~1(a). It is covered by sets of SiO$_2$ gratings (Fig.~1(b)), perpendicular to the ZnO c-axis, with periods $\Lambda$. The shift of the propagation constant by $2\pi/\Lambda$  allows a direct access to the polariton guided mode dispersion. The sample W2 is a half-microcavity without any grating. The sample geometries together with fabrication details are presented in the supplementary material \cite{suppl} and in \cite{Perez2016}. The common aspect of both samples is the presence of regular horizontal cracks, as one can see in the SEM image (Fig.~1(c)). They appear because of the mismatch of the lattice parameter and the thermal expansion coefficient between ZnO and ZnMgO. They are perpendicular to the c-axis  of the crystal and typically separated by 5 to 40 $\mu$m. Reflecting light, they induce a substantial confinement leading to the formation of a genuine horizontal cavity for the guided modes. Their presence plays a crucial role in the success of our observations compared with the previous experimental studies of polariton waveguides. 

\begin{figure}[tbp]
 \begin{center}
 \includegraphics[scale=0.6]{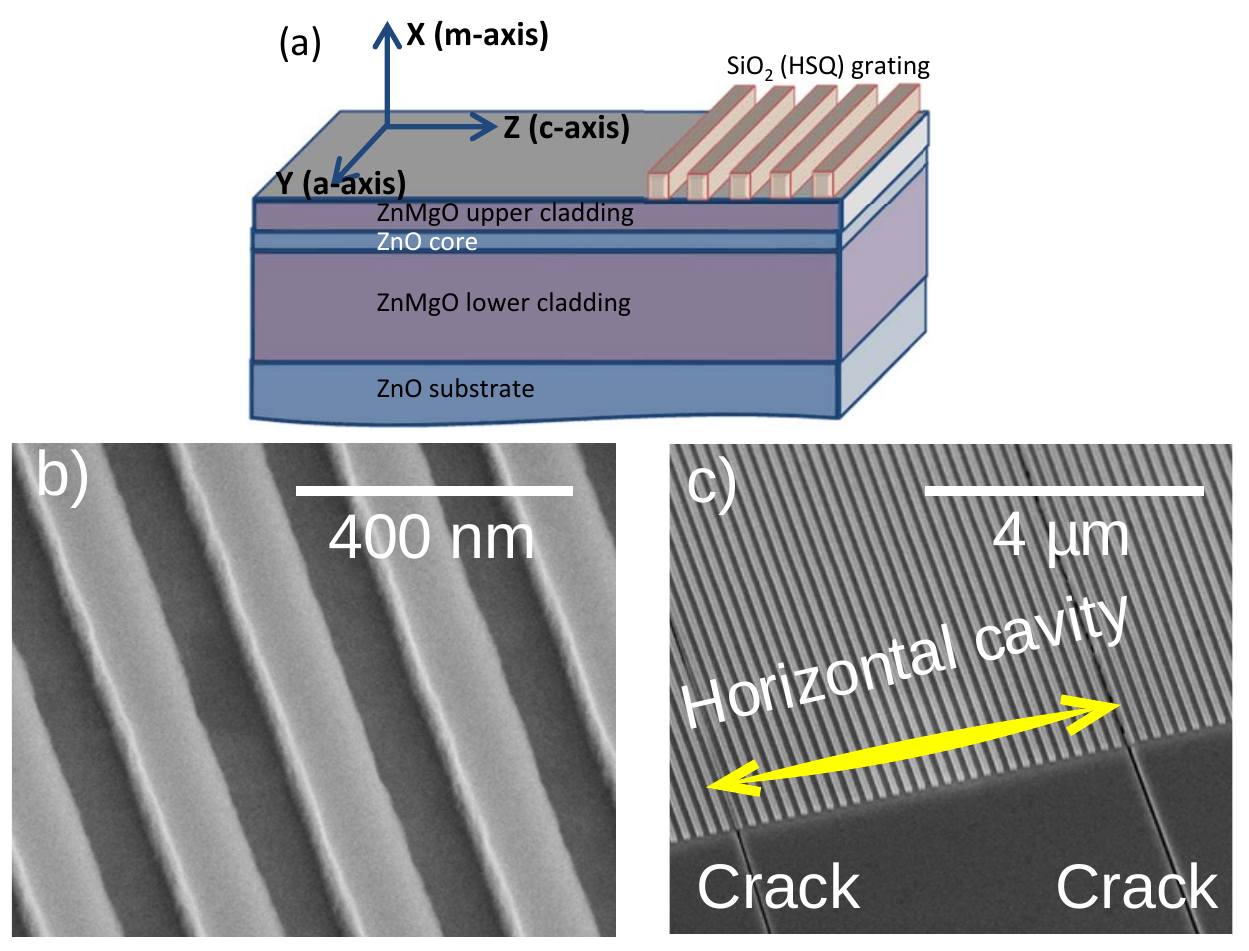}
 \caption{\label{fig1} Sample W1 description. (a): Scheme of the sample. The guided mode propagates along the Z-axis in the ZnO core. 
 (b): Scanning Electron Microscope (SEM) image of the SiO$_2$ grating deposited on top of the waveguide. 
(c): SEM image of sample showing two typical horizontal cracks here separated by 18 $\mu$m, nearly parallel to the grating.}
  \end{center}
 \end{figure}

The samples are studied through micro-photoluminescence using the fourth harmonic (266 nm) of a Nd:YAG laser as the excitation source. The pulse duration is 400 ps and the repetition rate is 20 kHz. A UV microscope objective with 0.4 numerical aperture was used to obtain small excitation spots of different diameters. The emission of the sample was collected through the same objective and imaged onto the spectrometer slit by a spherical lens mounted on a motorized translation stage.

The power dependencies of the emission of W1, at 5 and 300 K are shown in Fig.~2(a,b), demonstrating a very clear nonlinear threshold  at $P_{th}$. The horizontal cavity sizes are $~23 \mu$m and $~27 \mu$m at 5 K and 300 K  respectively. They are entirely covered by the pump spot and we do not use spatial selection of the emission, which originates from both the grating and the cracks. The guided polariton mode dispersions at 5 and 300 K, below and above threshold, are shown in Figs.~2(c,d) and 2(e,f), respectively. The optimal grating periods are $\Lambda=190 nm$ at 5K and $\Lambda=180 nm$ at 300K. All measured dispersions clearly deviate from the bare photonic mode (shown in dashed line together with the bare A-exciton energy). The main emission below threshold comes from donor-bound excitons (D$^0$X) located about 15 meV below the A-exciton energy. The emission from the cracks appears as a weak flat emission line,  which can be removed by spatially selecting the emission coming from the grating only (Fig.~S2 of \cite{suppl}). The measured polariton dispersion is well reproduced by a two-oscillator model where both A and B excitonic resonances are modeled as a single oscillator placed at the A-exciton energy. We take a linear photon dispersion with a slope fixed by the one of the measured dispersions around 3.1 eV. The extracted Rabi splitting is $\Omega_R=224$ meV at low pumping for both temperatures. At $P_{th}$, the dispersion remains essentially unchanged at 5 K ($\Omega_R=222$ meV). Polariton lasing takes place in modes having an exciton fraction of about 90 \%. A movie with emission images versus angle and energy  over a wide pumping range (0.1 to 20 $P_{th}$) is presented in the Supplemental material \cite{suppl}.  The increase of the pumping intensity induces a blue shift of the modes, as shown in Fig.~2(a,b) and Fig.~S3 of \cite{suppl}. It is due to the screening of the exciton oscillator strength, which is more significant at 300K. The fit gives $\Omega_R=180$ meV, 20 \% weaker than at low pumping (Fig. (2f)). Polariton lasing occurs in more photon-like states than at 5K, as theoretically expected \cite{Solnyshkov2014}, but still with a substantial 50-60 \% excitonic fraction. The velocity of these modes is about 40-50 \%  of that of the bare photonic mode of the waveguide, approximately 60 $\mu$m/ps. Despite the blue shift of the modes, the specific shape of the \emph{guided} polariton dispersion showing no energy minimum allows, when the pumping increases, to observe a red-shift of the emission intensity which nevertheless always remains on the slightly blueshifted polariton dispersion. This red-shift, clearly visible in the movie and in Fig.~S3 \cite{suppl}, and well reproduced by simulations based on semi-classical Boltzmann equations (Fig.~S4 \cite{suppl}), is due to the faster polariton relaxation rates at higher densities.

Interestingly, the lasing mechanism in ZnO-based systems, such as nanowires \cite{Huang2001}, has been a matter of debate for quite a long time. Polariton lasing \cite{Zamfirescu2002} was considered as a possible mechanism during a decade  \cite{Chu2008,Vanmaekelbergh2011,Versteegh2011,Versteegh2012}, but the absence of direct dispersion measurements made difficult to draw a definite conclusion. The direct dispersion measurements we report allow to clearly establish the polaritonic origin of the emission above the non-linear threshold.  The temperature dependences between 5 and 300 K of the polariton lasing energy and of the A-exciton energy are shown on Fig.~2(g). The difference between the two energies slightly increases, which demonstrates that lasing occurs in more and more photonic polariton states versus temperature. 
This behavior is expected from the previous numerical study of this system \cite{Solnyshkov2014} and from previous works performed on planar cavities\cite{Kasprzak2008,Levrat2010,Feng2013,Jamadi2016}. Indeed, similar to the rise of pumping, the increase of temperature leads to faster relaxation along the polariton branch which allows lasing to take place at lower polariton states.

\begin{figure}[tbp]
 \begin{center}
 \includegraphics[scale=0.42]{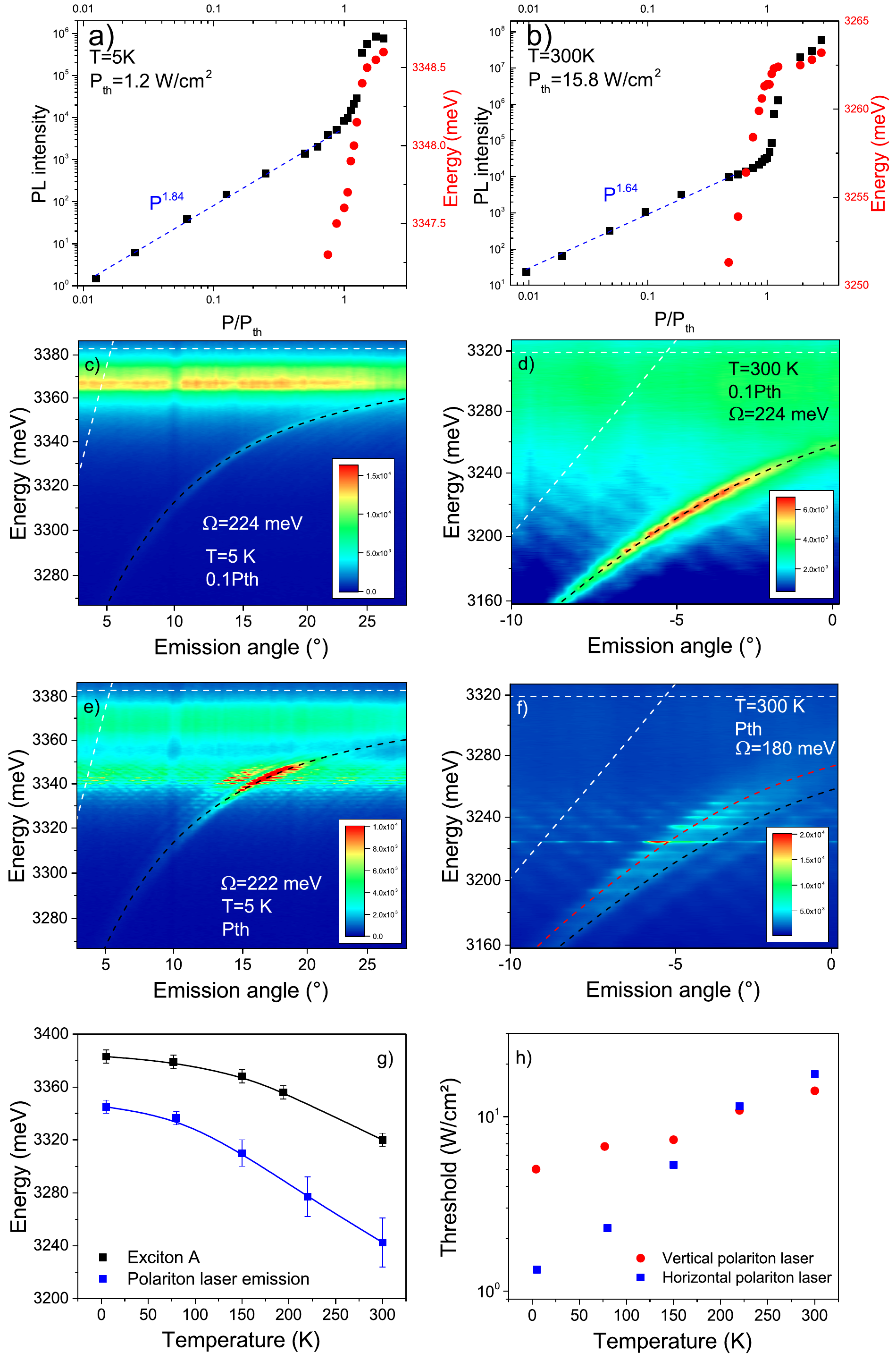}
 \caption{\label{fig2} Emission of W1 at 5 and 300K. The pump spot covers the entire horizontal cavity length. 
(a): 5K.  Power dependence of emission (black squares) and of the energy at the wave vector of the most intense lasing peaks.
(b): Same as (a), but at 300K.
(c-f): Energy of emission versus angle (dispersion). The white dashed lines are the expected bare TE0 photon and exciton modes. The black dashed lines are the strongly coupled polariton modes with a $\Omega_R$ indicated on each panel. The intensities indicated are in arbitrary units. 2(c): 5K and 0.1 $P_{th}$, 2(d): 300 K and 0.1 $P_{th}$, 2(e): 5K and $P_{th}$, 2(f) 300K and $P_{th}$. The red dashed line on 2(f) is the polariton dispersion with a $\Omega_R=224$ meV. (g): Experimental average exciton energies plotted together with the average horizontal polariton laser emission energies versus temperature. The solid black line corresponds to Vi\~na's law \cite{Vina1984} and the blue one is a guide for the eyes.
(h): Polariton lasing threshold versus temperature for a horizontal polariton laser in W1 (blue squares) and for a vertical cavity polariton laser studied in \cite{Jamadi2016} (red stars).
}
\end{center}
\end{figure}

Figure 2(h) shows a comparison of the thresholds measured in W1 (horizontal polariton lasing) and those of a vertical polariton laser measured in a full microcavity displaying a quality factor of 2000 \cite{Jamadi2016}, which corresponds to a cavity photon lifetime of 0.4 ps. Thresholds are comparable, being even slightly lower for the horizontal polariton laser in a wide temperature range, which can be understood qualitatively as follows. In both cases, the main polariton scattering mechanism is what was termed in the 60s "excitonic gain"  \cite{Haug1967}. It consists in the scattering of an exciton-like polariton either on another exciton-like polariton or on a LO-phonon towards a polariton state with lower energy and, thus, with a larger photonic fraction. In vertical cavities, polariton lasing takes place if excitonic gain is efficient enough compared with the polariton lifetime. In a guided mode geometry, the scattering rate should be compared with the transit time of polaritons under the pumping spot. Considering a typical 20 $\mu$m size pumping area and polariton guided modes propagating at 13 $\mu$m/ps at 5K and 60 $\mu$m/ps at 300 K, the transit time of polaritons under the spot is 1.5 ps and 0.33 ps respectively (neglecting the feedback provided by the horizontal cavities), quite comparable with the 0.4 ps of a  vertical ZnO-based cavity \cite{Jamadi2016}. This simple estimate explains the similarity of the measured thresholds in the two types of samples, and even the lower threshold  of the horizontal polariton laser at 5K. It also suggests that in thick microcavities, which support both radiative and guided modes, the latter are responsible for strong losses affecting vertical polariton lasing. In a cracked homoepitaxial full cavity (completed with a top dielectric DBR) we have been able to observe simultaneously lasing in the vertical radiative modes and in the horizontal guided modes (not shown).

Another interesting feature is the emergence of sharp emission lines, appearing at threshold within the polariton dispersion. These lines are the Fabry-Perot modes of the horizontal cavity formed by the two cracks surrounding the pump. The rise of these sharp peaks demonstrates the onset of phase coherence of the polariton modes all along the horizontal cavity, as  explained below.  
Figure 3(a) shows the real space emission at 5 K of the sample W2 excited above $P_{th}$ by a 7 $\mu$m size pump (4 times smaller than the distance between the surrounding cracks). The emission spectra from the pump area and from three different cracks are shown in Fig.~3(b). The pump area shows the radiative ZnO emission dominated by the first Bragg mode of the mirror at 3355 meV and the D$^0$X lines at 3370 meV. The free exciton line is visible as a shoulder around 3385 meV. The cracks' emission shows the Fabry-Perot interference peaks.  The power-dependent emission spectra measured at the crack "b",  demonstrating the emergence of the horizontal Fabry-Perot modes above threshold, are shown in Fig.~3(c). The interference fringes are weakly visible in the pump area since there is no well-defined out-coupling mechanism for the guided modes.  

\begin{figure}[H]
 \begin{center}
 \includegraphics[scale=0.33]{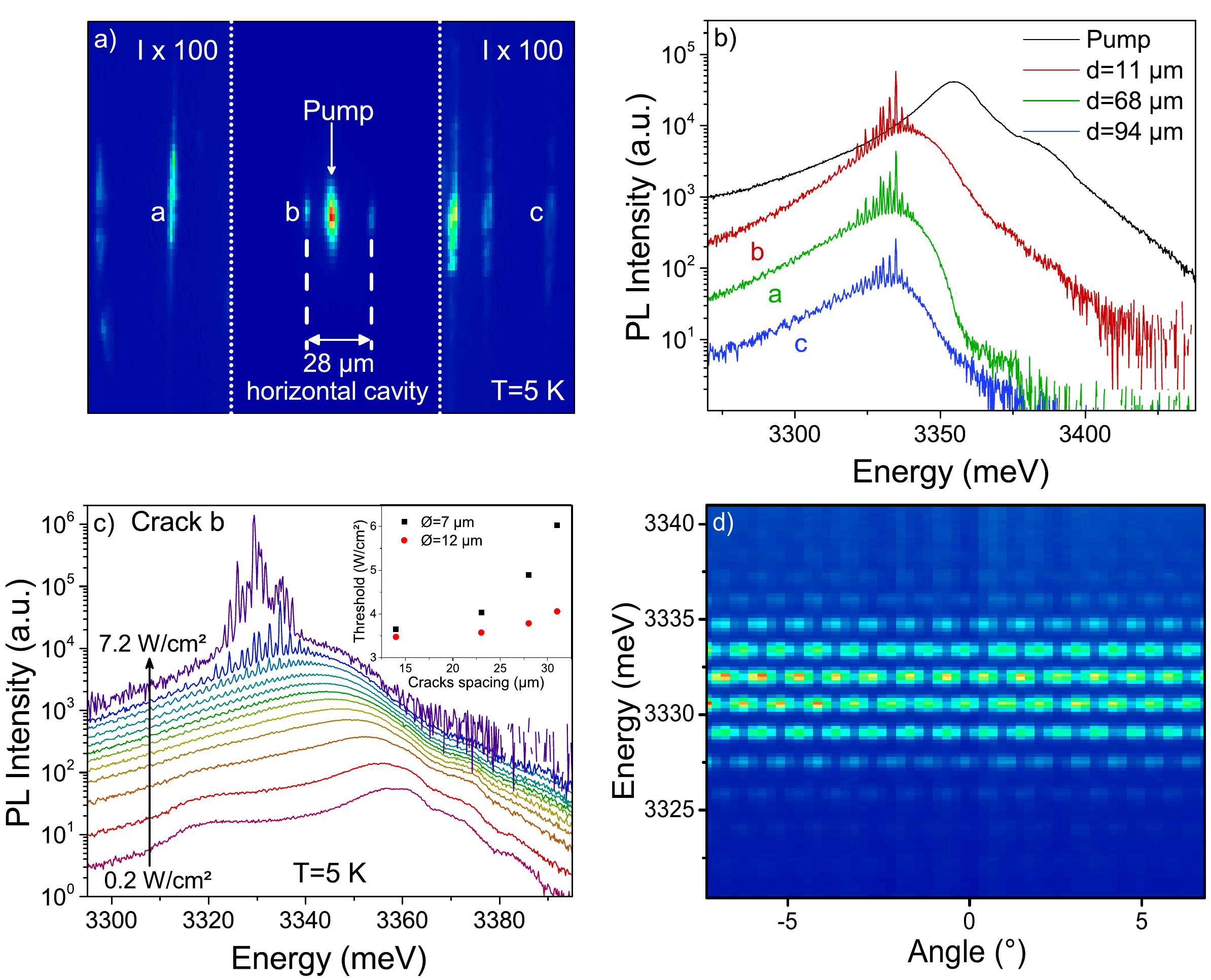}
 \caption{\label{fig3} Sample W2.
3(a): Real space emission at 5K. The intensity (I) of the two lateral regions is magnified by 100. 
3(b): Emission spectra from the pump region and from the cracks labeled "a", "b", and "c" in Fig.~3(a). $d$ is the distance between the crack and the center of the excitation spot. 
3(c): Emission of the crack "b" versus pumping power, showing the emergence of the Fabry-Perot modes of the horizontal cavity together with the onset of lasing (power dependence is shown in \cite{suppl}). The inset shows the threshold values versus the cavity size for two different pump diameters.
3(d): Far field image zoomed on the fringes resulting from the interference of the emission from two distant cracks demonstrating their mutual coherence.}
\end{center}
\end{figure}

Figure 3(d) is a far-field image resulting from the interference of the emission from two distant cracks, which clearly demonstrates their mutual coherence. The inset of Fig.~3(c) shows the influence of the spot size and of the distance between cracks on $P_{th}$. As one could expect, the smallest threshold is achieved when the pump and cavity size are comparable. However, the dependence is relatively weak and a pump about 4 times smaller than the cavity shows only a doubled threshold, because the residual absorption in the region without pumping is extremely reduced at the polariton energy. This is typical for the polariton lasing mechanism, which allows gain (polariton stimulated scattering) to take place well below the energy of electronic transitions.
Interestingly, the presence of the regular Fabry-Perot interference fringes for a known crack separation allows indirect extraction of the dispersion. This method is commonly used in the study of nanowires \cite{Vanmaekelbergh2011,Versteegh2012}, where direct dispersion measurements are difficult. The result of  extraction is shown in Fig.~S6 of \cite{suppl}. It reveals a curved dispersion typical for polaritons with $\Omega_R\sim 200$ meV.

\begin{figure}[tbp]
 \begin{center}
 \includegraphics[scale=0.5]{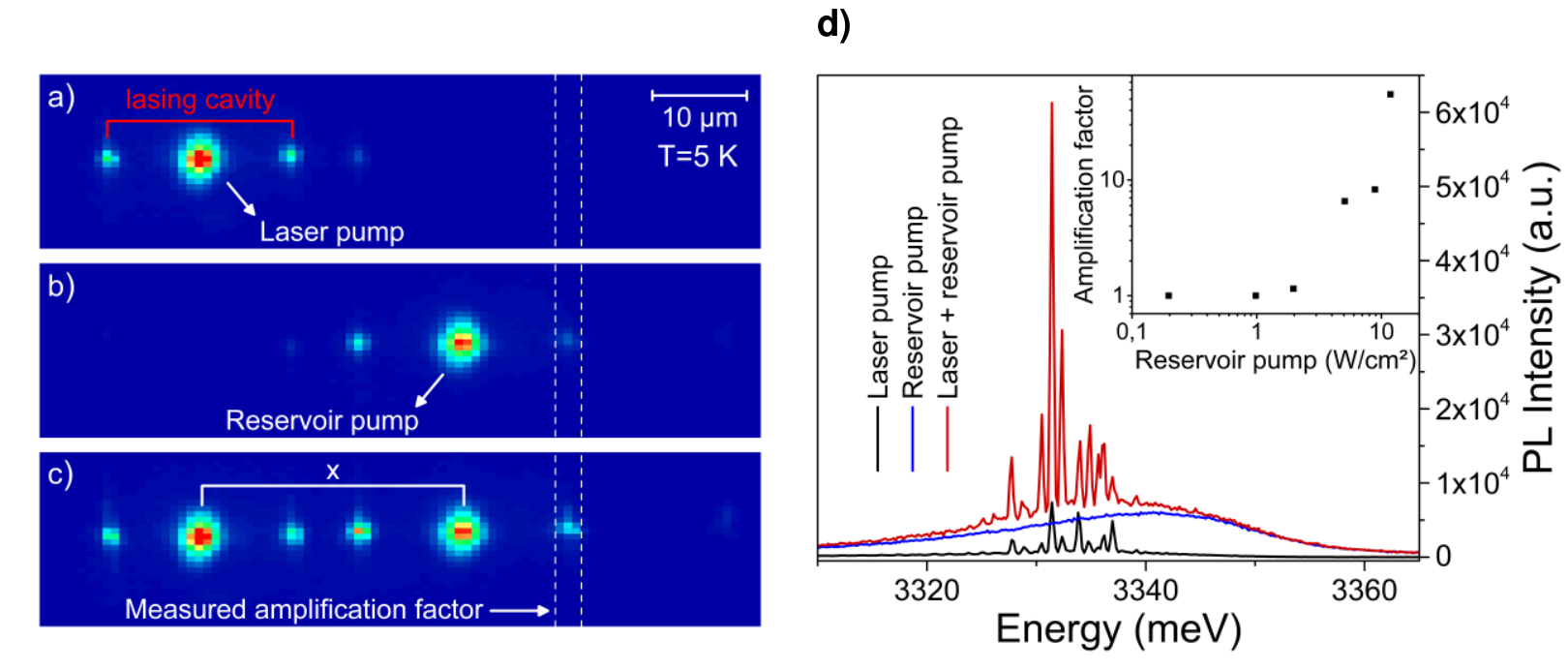}
 \caption{\label{fig4} Horizontal polariton amplifier regime at 5 K in W2.
(a,b,c): Real space emission with: (a) only "laser pump" (above lasing threshold); b) only "reservoir pump" (below lasing threshold); c) both pump spots ("laser" and "reservoir"). The vertical dashed lines show the crack where the amplification is measured.
4(d): Emission spectra extracted through the crack at the right of the reservoir pump (60 $\mu$m from the laser pump). The "reservoir" pump power is 8.9 W/cm$^2$, which is below threshold for this specific cavity as one can see from the emission spectrum when only the "reservoir pump" is ON. The "laser" pump power is 6.9 W/cm$^2$ -- above threshold for this cavity. The amplification factor is defined as the ratio between the most intense lasing peaks. The inset shows its evolution as a function of the reservoir pump.}
\end{center}
\end{figure}

Next, we demonstrate by a two-spot experiment how the horizontal polariton laser emission created in a given cavity can propagate and be reamplified several tens of $\mu$m away by another pump, being itself below the lasing threshold (Fig.~4). A first spot, named "laser pump", excites a horizontal cavity, creating the laser signal to be amplified (Fig.~4(a)). A second spot, named "reservoir pump", is placed in another horizontal cavity, at a distance of 60 $\mu$m from the laser pump (Fig.~4(b)). The power density of the laser pump is fixed above threshold, while the one of the reservoir pump is kept below threshold. Fig.~4(c) shows the configuration when both spots are turned ON. The emission spectra from the selected crack are displayed for each of the three previous pumping schemes in Fig.~4(d). The horizontal polariton laser signal is amplified when the two pumps are ON, with a maximum amplification factor of 55 (inset of Fig.~4(d)). This proves that the signal created by the laser pump, which can be easily identified by its free-spectral range defined by the initial cavity and which decays by about one order of magnitude when reaching the reservoir pump, is strongly reamplified. This can also be interpreted as an optical transistor, the transmission of a signal from a source (the laser pump) to a drain (the crack) being modulated by a control gate (the reservoir).

To summarize, we have demonstrated a horizontal polariton laser and amplifier, much simpler to fabricate and to process than full vertical microcavities, and simpler to manipulate than nanowires with which our system shares many similarities. One crucial step for polaritonics is electrical injection \cite{Bhattacharya2014,Schneider2013} which will be strongly facilitated in this horizontal geometry. Thus, the whole family of model devices implemented at low temperature in GaAs-based samples can be probably implemented at room temperature, under electrical injection, using alternative materials (ZnO, GaN, organics, TMD, ... ), and displaying greatly reduced photonic losses thanks to the guided modes. These exciting perspectives open the way to many studies in the future.

\begin{acknowledgments}
We acknowledge the support of the ANR projects: "Plug and Bose" (ANR-16-CE24-0021), "Quantum Fluids of Light" (ANR-16-CE30-0021) and the "Investissement d’avenir" program GANEX (ANR-11-LABX-004), IMOBS3 (ANR-10-LABX-16-01), ISITE "Cap2025". C2N is a member of RENATECH (CNRS), the national network of large micro-nanofabrication facilities. S.B. thank Dr F. Raineri of C2N for fruitful discussions. D.D.S. acknowledges the support of IUF (Institut Universitaire de France).
\end{acknowledgments}

\bibliography{reference}

\begin{thebibliography}{64}
\expandafter\ifx\csname natexlab\endcsname\relax\def\natexlab#1{#1}\fi
\expandafter\ifx\csname bibnamefont\endcsname\relax
  \def\bibnamefont#1{#1}\fi
\expandafter\ifx\csname bibfnamefont\endcsname\relax
  \def\bibfnamefont#1{#1}\fi
\expandafter\ifx\csname citenamefont\endcsname\relax
  \def\citenamefont#1{#1}\fi
\expandafter\ifx\csname url\endcsname\relax
  \def\url#1{\texttt{#1}}\fi
\expandafter\ifx\csname urlprefix\endcsname\relax\def\urlprefix{URL }\fi
\providecommand{\bibinfo}[2]{#2}
\providecommand{\eprint}[2][]{\url{#2}}

\bibitem[{\citenamefont{Hopfield}(1958)}]{Hopfield1958}
\bibinfo{author}{\bibfnamefont{J.~J.} \bibnamefont{Hopfield}},
  \bibinfo{journal}{Phys. Rev.} \textbf{\bibinfo{volume}{112}},
  \bibinfo{pages}{1555} (\bibinfo{year}{1958}),
  \urlprefix\url{https://link.aps.org/doi/10.1103/PhysRev.112.1555}.

\bibitem[{\citenamefont{Agranovich}(1959)}]{Agranovich1959}
\bibinfo{author}{\bibfnamefont{V.~M.} \bibnamefont{Agranovich}},
  \bibinfo{journal}{Zh. Eksp. Teor. Fiz.} \textbf{\bibinfo{volume}{37}},
  \bibinfo{pages}{1555} (\bibinfo{year}{1959}).

\bibitem[{\citenamefont{Weisbuch et~al.}(1992)\citenamefont{Weisbuch, Nishioka,
  Ishikawa, and Arakawa}}]{Weisbuch1992}
\bibinfo{author}{\bibfnamefont{C.}~\bibnamefont{Weisbuch}},
  \bibinfo{author}{\bibfnamefont{M.}~\bibnamefont{Nishioka}},
  \bibinfo{author}{\bibfnamefont{A.}~\bibnamefont{Ishikawa}}, \bibnamefont{and}
  \bibinfo{author}{\bibfnamefont{Y.}~\bibnamefont{Arakawa}},
  \bibinfo{journal}{Phys. Rev. Lett.} \textbf{\bibinfo{volume}{69}},
  \bibinfo{pages}{3314} (\bibinfo{year}{1992}),
  \urlprefix\url{https://link.aps.org/doi/10.1103/PhysRevLett.69.3314}.

\bibitem[{\citenamefont{Shelykh et~al.}(2010)\citenamefont{Shelykh, Kavokin,
  Rubo, Liew, and Malpuech}}]{Shelykh2010}
\bibinfo{author}{\bibfnamefont{I.~A.} \bibnamefont{Shelykh}},
  \bibinfo{author}{\bibfnamefont{A.~V.} \bibnamefont{Kavokin}},
  \bibinfo{author}{\bibfnamefont{Y.~G.} \bibnamefont{Rubo}},
  \bibinfo{author}{\bibfnamefont{T.~C.~H.} \bibnamefont{Liew}},
  \bibnamefont{and} \bibinfo{author}{\bibfnamefont{G.}~\bibnamefont{Malpuech}},
  \bibinfo{journal}{Semiconductor Science and Technology}
  \textbf{\bibinfo{volume}{25}}, \bibinfo{pages}{013001}
  (\bibinfo{year}{2010}),
  \urlprefix\url{http://stacks.iop.org/0268-1242/25/i=1/a=013001}.

\bibitem[{\citenamefont{Carusotto and Ciuti}(2013)}]{Carusotto2013}
\bibinfo{author}{\bibfnamefont{I.}~\bibnamefont{Carusotto}} \bibnamefont{and}
  \bibinfo{author}{\bibfnamefont{C.}~\bibnamefont{Ciuti}},
  \bibinfo{journal}{Rev. Mod. Phys.} \textbf{\bibinfo{volume}{85}},
  \bibinfo{pages}{299} (\bibinfo{year}{2013}),
  \urlprefix\url{http://link.aps.org/doi/10.1103/RevModPhys.85.299}.

\bibitem[{\citenamefont{Richard et~al.}(2005)\citenamefont{Richard, Kasprzak,
  Andr\'e, Romestain, Dang, Malpuech, and Kavokin}}]{Richard2005}
\bibinfo{author}{\bibfnamefont{M.}~\bibnamefont{Richard}},
  \bibinfo{author}{\bibfnamefont{J.}~\bibnamefont{Kasprzak}},
  \bibinfo{author}{\bibfnamefont{R.}~\bibnamefont{Andr\'e}},
  \bibinfo{author}{\bibfnamefont{R.}~\bibnamefont{Romestain}},
  \bibinfo{author}{\bibfnamefont{L.~S.} \bibnamefont{Dang}},
  \bibinfo{author}{\bibfnamefont{G.}~\bibnamefont{Malpuech}}, \bibnamefont{and}
  \bibinfo{author}{\bibfnamefont{A.}~\bibnamefont{Kavokin}},
  \bibinfo{journal}{Phys. Rev. B} \textbf{\bibinfo{volume}{72}},
  \bibinfo{pages}{201301} (\bibinfo{year}{2005}),
  \urlprefix\url{https://link.aps.org/doi/10.1103/PhysRevB.72.201301}.

\bibitem[{\citenamefont{Kasprzak et~al.}(2006)\citenamefont{Kasprzak, Richard,
  Kundermann, Baas, Jeambrun, Keeling, Marchetti, Szymanska, Andre, Staehli
  et~al.}}]{Kasprzak2006}
\bibinfo{author}{\bibfnamefont{J.}~\bibnamefont{Kasprzak}},
  \bibinfo{author}{\bibfnamefont{M.}~\bibnamefont{Richard}},
  \bibinfo{author}{\bibfnamefont{S.}~\bibnamefont{Kundermann}},
  \bibinfo{author}{\bibfnamefont{A.}~\bibnamefont{Baas}},
  \bibinfo{author}{\bibfnamefont{P.}~\bibnamefont{Jeambrun}},
  \bibinfo{author}{\bibfnamefont{J.~M.~J.} \bibnamefont{Keeling}},
  \bibinfo{author}{\bibfnamefont{F.~M.} \bibnamefont{Marchetti}},
  \bibinfo{author}{\bibfnamefont{M.~H.} \bibnamefont{Szymanska}},
  \bibinfo{author}{\bibfnamefont{R.}~\bibnamefont{Andre}},
  \bibinfo{author}{\bibfnamefont{J.~L.} \bibnamefont{Staehli}},
  \bibnamefont{et~al.}, \bibinfo{journal}{Nature}
  \textbf{\bibinfo{volume}{443}}, \bibinfo{pages}{409} (\bibinfo{year}{2006}),
  ISSN \bibinfo{issn}{0028-0836}.

\bibitem[{\citenamefont{Kasprzak et~al.}(2008)\citenamefont{Kasprzak,
  Solnyshkov, Andr\'e, Dang, and Malpuech}}]{Kasprzak2008}
\bibinfo{author}{\bibfnamefont{J.}~\bibnamefont{Kasprzak}},
  \bibinfo{author}{\bibfnamefont{D.~D.} \bibnamefont{Solnyshkov}},
  \bibinfo{author}{\bibfnamefont{R.}~\bibnamefont{Andr\'e}},
  \bibinfo{author}{\bibfnamefont{L.~S.} \bibnamefont{Dang}}, \bibnamefont{and}
  \bibinfo{author}{\bibfnamefont{G.}~\bibnamefont{Malpuech}},
  \bibinfo{journal}{Phys. Rev. Lett.} \textbf{\bibinfo{volume}{101}},
  \bibinfo{pages}{146404} (\bibinfo{year}{2008}),
  \urlprefix\url{https://link.aps.org/doi/10.1103/PhysRevLett.101.146404}.

\bibitem[{\citenamefont{Sun et~al.}(2017)\citenamefont{Sun, Wen, Yoon, Liu,
  Steger, Pfeiffer, West, Snoke, and Nelson}}]{Sun2017}
\bibinfo{author}{\bibfnamefont{Y.}~\bibnamefont{Sun}},
  \bibinfo{author}{\bibfnamefont{P.}~\bibnamefont{Wen}},
  \bibinfo{author}{\bibfnamefont{Y.}~\bibnamefont{Yoon}},
  \bibinfo{author}{\bibfnamefont{G.}~\bibnamefont{Liu}},
  \bibinfo{author}{\bibfnamefont{M.}~\bibnamefont{Steger}},
  \bibinfo{author}{\bibfnamefont{L.~N.} \bibnamefont{Pfeiffer}},
  \bibinfo{author}{\bibfnamefont{K.}~\bibnamefont{West}},
  \bibinfo{author}{\bibfnamefont{D.~W.} \bibnamefont{Snoke}}, \bibnamefont{and}
  \bibinfo{author}{\bibfnamefont{K.~A.} \bibnamefont{Nelson}},
  \bibinfo{journal}{Phys. Rev. Lett.} \textbf{\bibinfo{volume}{118}},
  \bibinfo{pages}{016602} (\bibinfo{year}{2017}),
  \urlprefix\url{https://link.aps.org/doi/10.1103/PhysRevLett.118.016602}.

\bibitem[{\citenamefont{Amo et~al.}(2009)\citenamefont{Amo, Lefrere, Pigeon,
  Adrados, Ciuti, Carusotto, Houdre, Giacobino, and Bramati}}]{Amo2009}
\bibinfo{author}{\bibfnamefont{A.}~\bibnamefont{Amo}},
  \bibinfo{author}{\bibfnamefont{J.}~\bibnamefont{Lefrere}},
  \bibinfo{author}{\bibfnamefont{S.}~\bibnamefont{Pigeon}},
  \bibinfo{author}{\bibfnamefont{C.}~\bibnamefont{Adrados}},
  \bibinfo{author}{\bibfnamefont{C.}~\bibnamefont{Ciuti}},
  \bibinfo{author}{\bibfnamefont{I.}~\bibnamefont{Carusotto}},
  \bibinfo{author}{\bibfnamefont{R.}~\bibnamefont{Houdre}},
  \bibinfo{author}{\bibfnamefont{E.}~\bibnamefont{Giacobino}},
  \bibnamefont{and} \bibinfo{author}{\bibfnamefont{A.}~\bibnamefont{Bramati}},
  \bibinfo{journal}{Nature Physics} \textbf{\bibinfo{volume}{5}},
  \bibinfo{pages}{805} (\bibinfo{year}{2009}).

\bibitem[{\citenamefont{Sanvitto and Kena-Cohen}(2016)}]{Sanvitto2016}
\bibinfo{author}{\bibfnamefont{D.}~\bibnamefont{Sanvitto}} \bibnamefont{and}
  \bibinfo{author}{\bibfnamefont{S.}~\bibnamefont{Kena-Cohen}},
  \bibinfo{journal}{Nature Materials} \textbf{\bibinfo{volume}{15}},
  \bibinfo{pages}{1061} (\bibinfo{year}{2016}).

\bibitem[{\citenamefont{Lagoudakis et~al.}(2008)\citenamefont{Lagoudakis,
  Wouters, Richard, Baas, Carusotto, Andre, Dang, and
  Deveaud-Pledran}}]{Lagoudakis2008}
\bibinfo{author}{\bibfnamefont{K.~G.} \bibnamefont{Lagoudakis}},
  \bibinfo{author}{\bibfnamefont{M.}~\bibnamefont{Wouters}},
  \bibinfo{author}{\bibfnamefont{M.}~\bibnamefont{Richard}},
  \bibinfo{author}{\bibfnamefont{A.}~\bibnamefont{Baas}},
  \bibinfo{author}{\bibfnamefont{I.}~\bibnamefont{Carusotto}},
  \bibinfo{author}{\bibfnamefont{R.}~\bibnamefont{Andre}},
  \bibinfo{author}{\bibfnamefont{L.~S.} \bibnamefont{Dang}}, \bibnamefont{and}
  \bibinfo{author}{\bibfnamefont{B.}~\bibnamefont{Deveaud-Pledran}},
  \bibinfo{journal}{Nature Physics} \textbf{\bibinfo{volume}{4}},
  \bibinfo{pages}{706} (\bibinfo{year}{2008}).

\bibitem[{\citenamefont{Lagoudakis et~al.}(2009)\citenamefont{Lagoudakis,
  Ostatnick\'y, Kavokin, Rubo, Andr\'e, and
  Deveaud-Pl�dran}}]{Lagoudakis2009}
\bibinfo{author}{\bibfnamefont{K.~G.} \bibnamefont{Lagoudakis}},
  \bibinfo{author}{\bibfnamefont{T.}~\bibnamefont{Ostatnick\'y}},
  \bibinfo{author}{\bibfnamefont{A.~V.} \bibnamefont{Kavokin}},
  \bibinfo{author}{\bibfnamefont{Y.~G.} \bibnamefont{Rubo}},
  \bibinfo{author}{\bibfnamefont{R.}~\bibnamefont{Andr\'e}}, \bibnamefont{and}
  \bibinfo{author}{\bibfnamefont{B.}~\bibnamefont{Deveaud-Pl�dran}},
  \bibinfo{journal}{Science} \textbf{\bibinfo{volume}{326}},
  \bibinfo{pages}{974} (\bibinfo{year}{2009}).

\bibitem[{\citenamefont{Hivet et~al.}(2012)\citenamefont{Hivet, Flayac,
  Solnyshkov, Tanese, Boulier, Andreoli, Giacobino, Bloch, Bramati, Malpuech
  et~al.}}]{Hivet}
\bibinfo{author}{\bibfnamefont{R.}~\bibnamefont{Hivet}},
  \bibinfo{author}{\bibfnamefont{H.}~\bibnamefont{Flayac}},
  \bibinfo{author}{\bibfnamefont{D.~D.} \bibnamefont{Solnyshkov}},
  \bibinfo{author}{\bibfnamefont{D.}~\bibnamefont{Tanese}},
  \bibinfo{author}{\bibfnamefont{T.}~\bibnamefont{Boulier}},
  \bibinfo{author}{\bibfnamefont{D.}~\bibnamefont{Andreoli}},
  \bibinfo{author}{\bibfnamefont{E.}~\bibnamefont{Giacobino}},
  \bibinfo{author}{\bibfnamefont{J.}~\bibnamefont{Bloch}},
  \bibinfo{author}{\bibfnamefont{A.}~\bibnamefont{Bramati}},
  \bibinfo{author}{\bibfnamefont{G.}~\bibnamefont{Malpuech}},
  \bibnamefont{et~al.}, \bibinfo{journal}{Nat Phys}
  \textbf{\bibinfo{volume}{8}}, \bibinfo{pages}{724} (\bibinfo{year}{2012}),
  ISSN \bibinfo{issn}{1745-2473},
  \urlprefix\url{http://dx.doi.org/10.1038/nphys2406}.

\bibitem[{\citenamefont{Kavokin et~al.}(2005)\citenamefont{Kavokin, Malpuech,
  and Glazov}}]{Kavokin2005}
\bibinfo{author}{\bibfnamefont{A.}~\bibnamefont{Kavokin}},
  \bibinfo{author}{\bibfnamefont{G.}~\bibnamefont{Malpuech}}, \bibnamefont{and}
  \bibinfo{author}{\bibfnamefont{M.}~\bibnamefont{Glazov}},
  \bibinfo{journal}{Phys. Rev. Lett.} \textbf{\bibinfo{volume}{95}},
  \bibinfo{pages}{136601} (\bibinfo{year}{2005}),
  \urlprefix\url{http://link.aps.org/doi/10.1103/PhysRevLett.95.136601}.

\bibitem[{\citenamefont{Leyder et~al.}(2007)\citenamefont{Leyder, Romanelli,
  Karr, Giacobino, Liew, Glazov, Kavokin, Malpuech, and Bramati}}]{Leyder2007}
\bibinfo{author}{\bibfnamefont{C.}~\bibnamefont{Leyder}},
  \bibinfo{author}{\bibfnamefont{M.}~\bibnamefont{Romanelli}},
  \bibinfo{author}{\bibfnamefont{J.~P.} \bibnamefont{Karr}},
  \bibinfo{author}{\bibfnamefont{E.}~\bibnamefont{Giacobino}},
  \bibinfo{author}{\bibfnamefont{T.~C.~H.} \bibnamefont{Liew}},
  \bibinfo{author}{\bibfnamefont{M.~M.} \bibnamefont{Glazov}},
  \bibinfo{author}{\bibfnamefont{A.~V.} \bibnamefont{Kavokin}},
  \bibinfo{author}{\bibfnamefont{G.}~\bibnamefont{Malpuech}}, \bibnamefont{and}
  \bibinfo{author}{\bibfnamefont{A.}~\bibnamefont{Bramati}},
  \bibinfo{journal}{Nat Phys} \textbf{\bibinfo{volume}{3}},
  \bibinfo{pages}{628} (\bibinfo{year}{2007}), ISSN \bibinfo{issn}{1745-2473},
  \urlprefix\url{http://dx.doi.org/10.1038/nphys676}.

\bibitem[{\citenamefont{Galbiati et~al.}(2012)\citenamefont{Galbiati, Ferrier,
  Solnyshkov, Tanese, Wertz, Amo, Abbarchi, Senellart, Sagnes, Lemaitre
  et~al.}}]{Galbiati2012}
\bibinfo{author}{\bibfnamefont{M.}~\bibnamefont{Galbiati}},
  \bibinfo{author}{\bibfnamefont{L.}~\bibnamefont{Ferrier}},
  \bibinfo{author}{\bibfnamefont{D.~D.} \bibnamefont{Solnyshkov}},
  \bibinfo{author}{\bibfnamefont{D.}~\bibnamefont{Tanese}},
  \bibinfo{author}{\bibfnamefont{E.}~\bibnamefont{Wertz}},
  \bibinfo{author}{\bibfnamefont{A.}~\bibnamefont{Amo}},
  \bibinfo{author}{\bibfnamefont{M.}~\bibnamefont{Abbarchi}},
  \bibinfo{author}{\bibfnamefont{P.}~\bibnamefont{Senellart}},
  \bibinfo{author}{\bibfnamefont{I.}~\bibnamefont{Sagnes}},
  \bibinfo{author}{\bibfnamefont{A.}~\bibnamefont{Lemaitre}},
  \bibnamefont{et~al.}, \bibinfo{journal}{Phys. Rev. Lett.}
  \textbf{\bibinfo{volume}{108}}, \bibinfo{pages}{126403}
  (\bibinfo{year}{2012}).

\bibitem[{\citenamefont{Sala et~al.}(2015)\citenamefont{Sala, Solnyshkov,
  Carusotto, Jacqmin, Lema\^{\i}tre, Ter\ifmmode~\mbox{\c{c}}\else
  \c{c}\fi{}as, Nalitov, Abbarchi, Galopin, Sagnes et~al.}}]{Sala2014}
\bibinfo{author}{\bibfnamefont{V.~G.} \bibnamefont{Sala}},
  \bibinfo{author}{\bibfnamefont{D.~D.} \bibnamefont{Solnyshkov}},
  \bibinfo{author}{\bibfnamefont{I.}~\bibnamefont{Carusotto}},
  \bibinfo{author}{\bibfnamefont{T.}~\bibnamefont{Jacqmin}},
  \bibinfo{author}{\bibfnamefont{A.}~\bibnamefont{Lema\^{\i}tre}},
  \bibinfo{author}{\bibfnamefont{H.}~\bibnamefont{Ter\ifmmode~\mbox{\c{c}}\else
  \c{c}\fi{}as}}, \bibinfo{author}{\bibfnamefont{A.}~\bibnamefont{Nalitov}},
  \bibinfo{author}{\bibfnamefont{M.}~\bibnamefont{Abbarchi}},
  \bibinfo{author}{\bibfnamefont{E.}~\bibnamefont{Galopin}},
  \bibinfo{author}{\bibfnamefont{I.}~\bibnamefont{Sagnes}},
  \bibnamefont{et~al.}, \bibinfo{journal}{Phys. Rev. X}
  \textbf{\bibinfo{volume}{5}}, \bibinfo{pages}{011034} (\bibinfo{year}{2015}),
  \urlprefix\url{http://link.aps.org/doi/10.1103/PhysRevX.5.011034}.

\bibitem[{\citenamefont{Lai et~al.}(2007)\citenamefont{Lai, Kim, Utsunomiya,
  Roumpos, Deng, Fraser, Byrnes, Recher, Kumada, Fujisawa et~al.}}]{Lai2007}
\bibinfo{author}{\bibfnamefont{C.~W.} \bibnamefont{Lai}},
  \bibinfo{author}{\bibfnamefont{N.~Y.} \bibnamefont{Kim}},
  \bibinfo{author}{\bibfnamefont{S.}~\bibnamefont{Utsunomiya}},
  \bibinfo{author}{\bibfnamefont{G.}~\bibnamefont{Roumpos}},
  \bibinfo{author}{\bibfnamefont{H.}~\bibnamefont{Deng}},
  \bibinfo{author}{\bibfnamefont{M.~D.} \bibnamefont{Fraser}},
  \bibinfo{author}{\bibfnamefont{T.}~\bibnamefont{Byrnes}},
  \bibinfo{author}{\bibfnamefont{P.}~\bibnamefont{Recher}},
  \bibinfo{author}{\bibfnamefont{N.}~\bibnamefont{Kumada}},
  \bibinfo{author}{\bibfnamefont{T.}~\bibnamefont{Fujisawa}},
  \bibnamefont{et~al.}, \bibinfo{journal}{Nature}
  \textbf{\bibinfo{volume}{450}}, \bibinfo{pages}{529} (\bibinfo{year}{2007}).

\bibitem[{\citenamefont{Jacqmin et~al.}(2014)\citenamefont{Jacqmin, Carusotto,
  Sagnes, Abbarchi, Solnyshkov, Malpuech, Galopin, Lemaitre, Bloch, and
  Amo}}]{Jacqmin2014}
\bibinfo{author}{\bibfnamefont{T.}~\bibnamefont{Jacqmin}},
  \bibinfo{author}{\bibfnamefont{I.}~\bibnamefont{Carusotto}},
  \bibinfo{author}{\bibfnamefont{I.}~\bibnamefont{Sagnes}},
  \bibinfo{author}{\bibfnamefont{M.}~\bibnamefont{Abbarchi}},
  \bibinfo{author}{\bibfnamefont{D.}~\bibnamefont{Solnyshkov}},
  \bibinfo{author}{\bibfnamefont{G.}~\bibnamefont{Malpuech}},
  \bibinfo{author}{\bibfnamefont{E.}~\bibnamefont{Galopin}},
  \bibinfo{author}{\bibfnamefont{A.}~\bibnamefont{Lemaitre}},
  \bibinfo{author}{\bibfnamefont{J.}~\bibnamefont{Bloch}}, \bibnamefont{and}
  \bibinfo{author}{\bibfnamefont{A.}~\bibnamefont{Amo}},
  \bibinfo{journal}{Phys. Rev. Lett.} \textbf{\bibinfo{volume}{112}},
  \bibinfo{pages}{116402} (\bibinfo{year}{2014}),
  \urlprefix\url{http://link.aps.org/doi/10.1103/PhysRevLett.112.116402}.

\bibitem[{\citenamefont{Kim et~al.}(2014)\citenamefont{Kim, Kusudo, L\"offler,
  H\"ofling, Forchel, and Yamamoto}}]{Kim2014}
\bibinfo{author}{\bibfnamefont{N.~Y.} \bibnamefont{Kim}},
  \bibinfo{author}{\bibfnamefont{K.}~\bibnamefont{Kusudo}},
  \bibinfo{author}{\bibfnamefont{A.}~\bibnamefont{L\"offler}},
  \bibinfo{author}{\bibfnamefont{S.}~\bibnamefont{H\"ofling}},
  \bibinfo{author}{\bibfnamefont{A.}~\bibnamefont{Forchel}}, \bibnamefont{and}
  \bibinfo{author}{\bibfnamefont{Y.}~\bibnamefont{Yamamoto}},
  \bibinfo{journal}{Phys. Rev. B} \textbf{\bibinfo{volume}{89}},
  \bibinfo{pages}{085306} (\bibinfo{year}{2014}),
  \urlprefix\url{https://link.aps.org/doi/10.1103/PhysRevB.89.085306}.

\bibitem[{\citenamefont{Whittaker et~al.}(2017)\citenamefont{Whittaker,
  Cancellieri, Walker, Gulevich, Schomerus, Vaitiekus, Royall, Whittaker,
  Clarke, Iorsh et~al.}}]{Whittaker2017}
\bibinfo{author}{\bibfnamefont{C.~E.} \bibnamefont{Whittaker}},
  \bibinfo{author}{\bibfnamefont{E.}~\bibnamefont{Cancellieri}},
  \bibinfo{author}{\bibfnamefont{P.~M.} \bibnamefont{Walker}},
  \bibinfo{author}{\bibfnamefont{D.~R.} \bibnamefont{Gulevich}},
  \bibinfo{author}{\bibfnamefont{H.}~\bibnamefont{Schomerus}},
  \bibinfo{author}{\bibfnamefont{D.}~\bibnamefont{Vaitiekus}},
  \bibinfo{author}{\bibfnamefont{B.}~\bibnamefont{Royall}},
  \bibinfo{author}{\bibfnamefont{D.~M.} \bibnamefont{Whittaker}},
  \bibinfo{author}{\bibfnamefont{E.}~\bibnamefont{Clarke}},
  \bibinfo{author}{\bibfnamefont{I.~V.} \bibnamefont{Iorsh}},
  \bibnamefont{et~al.}, \bibinfo{journal}{arXiv:1705.03006}
  (\bibinfo{year}{2017}).

\bibitem[{\citenamefont{Nalitov et~al.}(2015)\citenamefont{Nalitov, Solnyshkov,
  and Malpuech}}]{Nalitov2014b}
\bibinfo{author}{\bibfnamefont{A.~V.} \bibnamefont{Nalitov}},
  \bibinfo{author}{\bibfnamefont{D.~D.} \bibnamefont{Solnyshkov}},
  \bibnamefont{and} \bibinfo{author}{\bibfnamefont{G.}~\bibnamefont{Malpuech}},
  \bibinfo{journal}{Phys. Rev. Lett.} \textbf{\bibinfo{volume}{114}},
  \bibinfo{pages}{116401} (\bibinfo{year}{2015}),
  \urlprefix\url{http://link.aps.org/doi/10.1103/PhysRevLett.114.116401}.

\bibitem[{\citenamefont{Bleu et~al.}(2017)\citenamefont{Bleu, Malpuech, and
  Solnyshkov}}]{Bleu2017z}
\bibinfo{author}{\bibfnamefont{O.}~\bibnamefont{Bleu}},
  \bibinfo{author}{\bibfnamefont{G.}~\bibnamefont{Malpuech}}, \bibnamefont{and}
  \bibinfo{author}{\bibfnamefont{D.~D.} \bibnamefont{Solnyshkov}},
  \bibinfo{journal}{arXiv:1709.01830}  (\bibinfo{year}{2017}).

\bibitem[{\citenamefont{Berloff et~al.}(2017)\citenamefont{Berloff, Silva,
  Kalinin, Askitopoulos, Topfer, Cilibrizzi, Langbein, and
  Lagoudakis}}]{Berloff2017}
\bibinfo{author}{\bibfnamefont{N.~G.} \bibnamefont{Berloff}},
  \bibinfo{author}{\bibfnamefont{M.}~\bibnamefont{Silva}},
  \bibinfo{author}{\bibfnamefont{K.}~\bibnamefont{Kalinin}},
  \bibinfo{author}{\bibfnamefont{A.}~\bibnamefont{Askitopoulos}},
  \bibinfo{author}{\bibfnamefont{J.~D.} \bibnamefont{Topfer}},
  \bibinfo{author}{\bibfnamefont{P.}~\bibnamefont{Cilibrizzi}},
  \bibinfo{author}{\bibfnamefont{W.}~\bibnamefont{Langbein}}, \bibnamefont{and}
  \bibinfo{author}{\bibfnamefont{P.~G.} \bibnamefont{Lagoudakis}},
  \bibinfo{journal}{Nature Materials}  (\bibinfo{year}{2017}).

\bibitem[{\citenamefont{Caulfield and Dolev}(2010)}]{Caulfield2010}
\bibinfo{author}{\bibfnamefont{H.~J.} \bibnamefont{Caulfield}}
  \bibnamefont{and} \bibinfo{author}{\bibfnamefont{S.}~\bibnamefont{Dolev}},
  \bibinfo{journal}{Nature Photonics} \textbf{\bibinfo{volume}{4}},
  \bibinfo{pages}{261} (\bibinfo{year}{2010}).

\bibitem[{\citenamefont{Imamoglu et~al.}(1996)\citenamefont{Imamoglu, Ram, Pau,
  and Yamamoto}}]{Imamoglu1996}
\bibinfo{author}{\bibfnamefont{A.}~\bibnamefont{Imamoglu}},
  \bibinfo{author}{\bibfnamefont{R.~J.} \bibnamefont{Ram}},
  \bibinfo{author}{\bibfnamefont{S.}~\bibnamefont{Pau}}, \bibnamefont{and}
  \bibinfo{author}{\bibfnamefont{Y.}~\bibnamefont{Yamamoto}},
  \bibinfo{journal}{Phys. Rev. A} \textbf{\bibinfo{volume}{53}},
  \bibinfo{pages}{4250} (\bibinfo{year}{1996}),
  \urlprefix\url{https://link.aps.org/doi/10.1103/PhysRevA.53.4250}.

\bibitem[{\citenamefont{Amo et~al.}(2010)\citenamefont{Amo, Liew, Adrados,
  Houdre, Giacobino, Kavokin, and Bramati}}]{Amo2010}
\bibinfo{author}{\bibfnamefont{A.}~\bibnamefont{Amo}},
  \bibinfo{author}{\bibfnamefont{T.~C.~H.} \bibnamefont{Liew}},
  \bibinfo{author}{\bibfnamefont{C.}~\bibnamefont{Adrados}},
  \bibinfo{author}{\bibfnamefont{R.}~\bibnamefont{Houdre}},
  \bibinfo{author}{\bibfnamefont{E.}~\bibnamefont{Giacobino}},
  \bibinfo{author}{\bibfnamefont{A.~V.} \bibnamefont{Kavokin}},
  \bibnamefont{and} \bibinfo{author}{\bibfnamefont{A.}~\bibnamefont{Bramati}},
  \bibinfo{journal}{Nature Photonics} \textbf{\bibinfo{volume}{4}},
  \bibinfo{pages}{361} (\bibinfo{year}{2010}).

\bibitem[{\citenamefont{Gao et~al.}(2012)\citenamefont{Gao, Eldridge, Liew,
  Tsintzos, Stavrinidis, Deligeorgis, Hatzopoulos, and Savvidis}}]{Gao2012}
\bibinfo{author}{\bibfnamefont{T.}~\bibnamefont{Gao}},
  \bibinfo{author}{\bibfnamefont{P.~S.} \bibnamefont{Eldridge}},
  \bibinfo{author}{\bibfnamefont{T.~C.~H.} \bibnamefont{Liew}},
  \bibinfo{author}{\bibfnamefont{S.~I.} \bibnamefont{Tsintzos}},
  \bibinfo{author}{\bibfnamefont{G.}~\bibnamefont{Stavrinidis}},
  \bibinfo{author}{\bibfnamefont{G.}~\bibnamefont{Deligeorgis}},
  \bibinfo{author}{\bibfnamefont{Z.}~\bibnamefont{Hatzopoulos}},
  \bibnamefont{and} \bibinfo{author}{\bibfnamefont{P.~G.}
  \bibnamefont{Savvidis}}, \bibinfo{journal}{Phys. Rev. B}
  \textbf{\bibinfo{volume}{85}}, \bibinfo{pages}{235102}
  (\bibinfo{year}{2012}),
  \urlprefix\url{https://link.aps.org/doi/10.1103/PhysRevB.85.235102}.

\bibitem[{\citenamefont{Wertz et~al.}(2012)\citenamefont{Wertz, Amo,
  Solnyshkov, Ferrier, Liew, Sanvitto, Senellart, Sagnes, Lema\^{\i}tre,
  Kavokin et~al.}}]{Wertz2012}
\bibinfo{author}{\bibfnamefont{E.}~\bibnamefont{Wertz}},
  \bibinfo{author}{\bibfnamefont{A.}~\bibnamefont{Amo}},
  \bibinfo{author}{\bibfnamefont{D.~D.} \bibnamefont{Solnyshkov}},
  \bibinfo{author}{\bibfnamefont{L.}~\bibnamefont{Ferrier}},
  \bibinfo{author}{\bibfnamefont{T.~C.~H.} \bibnamefont{Liew}},
  \bibinfo{author}{\bibfnamefont{D.}~\bibnamefont{Sanvitto}},
  \bibinfo{author}{\bibfnamefont{P.}~\bibnamefont{Senellart}},
  \bibinfo{author}{\bibfnamefont{I.}~\bibnamefont{Sagnes}},
  \bibinfo{author}{\bibfnamefont{A.}~\bibnamefont{Lema\^{\i}tre}},
  \bibinfo{author}{\bibfnamefont{A.~V.} \bibnamefont{Kavokin}},
  \bibnamefont{et~al.}, \bibinfo{journal}{Phys. Rev. Lett.}
  \textbf{\bibinfo{volume}{109}}, \bibinfo{pages}{216404}
  (\bibinfo{year}{2012}),
  \urlprefix\url{https://link.aps.org/doi/10.1103/PhysRevLett.109.216404}.

\bibitem[{\citenamefont{Ballarini et~al.}(2013)\citenamefont{Ballarini,
  De~Giorgi, Cancellieri, Houdre, Giacobino, Cingolani, Bramati, Gigli, and
  Sanvitto}}]{Ballarini2013}
\bibinfo{author}{\bibfnamefont{D.}~\bibnamefont{Ballarini}},
  \bibinfo{author}{\bibfnamefont{M.}~\bibnamefont{De~Giorgi}},
  \bibinfo{author}{\bibfnamefont{E.}~\bibnamefont{Cancellieri}},
  \bibinfo{author}{\bibfnamefont{R.}~\bibnamefont{Houdre}},
  \bibinfo{author}{\bibfnamefont{E.}~\bibnamefont{Giacobino}},
  \bibinfo{author}{\bibfnamefont{R.}~\bibnamefont{Cingolani}},
  \bibinfo{author}{\bibfnamefont{A.}~\bibnamefont{Bramati}},
  \bibinfo{author}{\bibfnamefont{G.}~\bibnamefont{Gigli}}, \bibnamefont{and}
  \bibinfo{author}{\bibfnamefont{D.}~\bibnamefont{Sanvitto}},
  \bibinfo{journal}{Nature Communications} \textbf{\bibinfo{volume}{4}},
  \bibinfo{pages}{1778} (\bibinfo{year}{2013}).

\bibitem[{\citenamefont{Nguyen et~al.}(2013)\citenamefont{Nguyen, Vishnevsky,
  Sturm, Tanese, Solnyshkov, Galopin, Lema\^{\i}tre, Sagnes, Amo, Malpuech
  et~al.}}]{Nguyen2013}
\bibinfo{author}{\bibfnamefont{H.~S.} \bibnamefont{Nguyen}},
  \bibinfo{author}{\bibfnamefont{D.}~\bibnamefont{Vishnevsky}},
  \bibinfo{author}{\bibfnamefont{C.}~\bibnamefont{Sturm}},
  \bibinfo{author}{\bibfnamefont{D.}~\bibnamefont{Tanese}},
  \bibinfo{author}{\bibfnamefont{D.}~\bibnamefont{Solnyshkov}},
  \bibinfo{author}{\bibfnamefont{E.}~\bibnamefont{Galopin}},
  \bibinfo{author}{\bibfnamefont{A.}~\bibnamefont{Lema\^{\i}tre}},
  \bibinfo{author}{\bibfnamefont{I.}~\bibnamefont{Sagnes}},
  \bibinfo{author}{\bibfnamefont{A.}~\bibnamefont{Amo}},
  \bibinfo{author}{\bibfnamefont{G.}~\bibnamefont{Malpuech}},
  \bibnamefont{et~al.}, \bibinfo{journal}{Phys. Rev. Lett.}
  \textbf{\bibinfo{volume}{110}}, \bibinfo{pages}{236601}
  (\bibinfo{year}{2013}),
  \urlprefix\url{https://link.aps.org/doi/10.1103/PhysRevLett.110.236601}.

\bibitem[{\citenamefont{Christopoulos et~al.}(2007)\citenamefont{Christopoulos,
  von H\"ogersthal, Grundy, Lagoudakis, Kavokin, Baumberg, Christmann, Butt\'e,
  Feltin, Carlin et~al.}}]{Christopoulos2007}
\bibinfo{author}{\bibfnamefont{S.}~\bibnamefont{Christopoulos}},
  \bibinfo{author}{\bibfnamefont{G.~B.~H.} \bibnamefont{von H\"ogersthal}},
  \bibinfo{author}{\bibfnamefont{A.~J.~D.} \bibnamefont{Grundy}},
  \bibinfo{author}{\bibfnamefont{P.~G.} \bibnamefont{Lagoudakis}},
  \bibinfo{author}{\bibfnamefont{A.~V.} \bibnamefont{Kavokin}},
  \bibinfo{author}{\bibfnamefont{J.~J.} \bibnamefont{Baumberg}},
  \bibinfo{author}{\bibfnamefont{G.}~\bibnamefont{Christmann}},
  \bibinfo{author}{\bibfnamefont{R.}~\bibnamefont{Butt\'e}},
  \bibinfo{author}{\bibfnamefont{E.}~\bibnamefont{Feltin}},
  \bibinfo{author}{\bibfnamefont{J.-F.} \bibnamefont{Carlin}},
  \bibnamefont{et~al.}, \bibinfo{journal}{Phys. Rev. Lett.}
  \textbf{\bibinfo{volume}{98}}, \bibinfo{pages}{126405}
  (\bibinfo{year}{2007}),
  \urlprefix\url{https://link.aps.org/doi/10.1103/PhysRevLett.98.126405}.

\bibitem[{\citenamefont{Christmann et~al.}(2008)\citenamefont{Christmann,
  Butte, Feltin, Carlin, and Grandjean}}]{Christmann2008}
\bibinfo{author}{\bibfnamefont{G.}~\bibnamefont{Christmann}},
  \bibinfo{author}{\bibfnamefont{R.}~\bibnamefont{Butte}},
  \bibinfo{author}{\bibfnamefont{E.}~\bibnamefont{Feltin}},
  \bibinfo{author}{\bibfnamefont{J.-F.} \bibnamefont{Carlin}},
  \bibnamefont{and}
  \bibinfo{author}{\bibfnamefont{N.}~\bibnamefont{Grandjean}},
  \bibinfo{journal}{Appl. Phys. Lett.} \textbf{\bibinfo{volume}{93}},
  \bibinfo{pages}{51102} (\bibinfo{year}{2008}).

\bibitem[{\citenamefont{Bhattacharya et~al.}(2014)\citenamefont{Bhattacharya,
  Frost, Deshpande, Baten, Hazari, and Das}}]{Bhattacharya2014}
\bibinfo{author}{\bibfnamefont{P.}~\bibnamefont{Bhattacharya}},
  \bibinfo{author}{\bibfnamefont{T.}~\bibnamefont{Frost}},
  \bibinfo{author}{\bibfnamefont{S.}~\bibnamefont{Deshpande}},
  \bibinfo{author}{\bibfnamefont{M.~Z.} \bibnamefont{Baten}},
  \bibinfo{author}{\bibfnamefont{A.}~\bibnamefont{Hazari}}, \bibnamefont{and}
  \bibinfo{author}{\bibfnamefont{A.}~\bibnamefont{Das}},
  \bibinfo{journal}{Phys. Rev. Lett.} \textbf{\bibinfo{volume}{112}},
  \bibinfo{pages}{236802} (\bibinfo{year}{2014}),
  \urlprefix\url{https://link.aps.org/doi/10.1103/PhysRevLett.112.236802}.

\bibitem[{\citenamefont{Li et~al.}(2013)\citenamefont{Li, Orosz, Kamoun,
  Bouchoule, Brimont, Disseix, Guillet, Lafosse, Leroux, Leymarie
  et~al.}}]{Feng2013}
\bibinfo{author}{\bibfnamefont{F.}~\bibnamefont{Li}},
  \bibinfo{author}{\bibfnamefont{L.}~\bibnamefont{Orosz}},
  \bibinfo{author}{\bibfnamefont{O.}~\bibnamefont{Kamoun}},
  \bibinfo{author}{\bibfnamefont{S.}~\bibnamefont{Bouchoule}},
  \bibinfo{author}{\bibfnamefont{C.}~\bibnamefont{Brimont}},
  \bibinfo{author}{\bibfnamefont{P.}~\bibnamefont{Disseix}},
  \bibinfo{author}{\bibfnamefont{T.}~\bibnamefont{Guillet}},
  \bibinfo{author}{\bibfnamefont{X.}~\bibnamefont{Lafosse}},
  \bibinfo{author}{\bibfnamefont{M.}~\bibnamefont{Leroux}},
  \bibinfo{author}{\bibfnamefont{J.}~\bibnamefont{Leymarie}},
  \bibnamefont{et~al.}, \bibinfo{journal}{Phys. Rev. Lett.}
  \textbf{\bibinfo{volume}{110}}, \bibinfo{pages}{196406}
  (\bibinfo{year}{2013}),
  \urlprefix\url{https://link.aps.org/doi/10.1103/PhysRevLett.110.196406}.

\bibitem[{\citenamefont{Kena-Cohen and Forrest}(2010)}]{Cohen2010}
\bibinfo{author}{\bibfnamefont{S.}~\bibnamefont{Kena-Cohen}} \bibnamefont{and}
  \bibinfo{author}{\bibfnamefont{S.~R.} \bibnamefont{Forrest}},
  \bibinfo{journal}{Nature Photonics} \textbf{\bibinfo{volume}{4}},
  \bibinfo{pages}{371} (\bibinfo{year}{2010}).

\bibitem[{\citenamefont{Dietrich et~al.}(2016)\citenamefont{Dietrich, Steude,
  Tropf, Schubert, Kronenberg, Ostermann, Hofling, and Gather}}]{Dietrich2016}
\bibinfo{author}{\bibfnamefont{C.~P.} \bibnamefont{Dietrich}},
  \bibinfo{author}{\bibfnamefont{A.}~\bibnamefont{Steude}},
  \bibinfo{author}{\bibfnamefont{L.}~\bibnamefont{Tropf}},
  \bibinfo{author}{\bibfnamefont{M.}~\bibnamefont{Schubert}},
  \bibinfo{author}{\bibfnamefont{N.~M.} \bibnamefont{Kronenberg}},
  \bibinfo{author}{\bibfnamefont{K.}~\bibnamefont{Ostermann}},
  \bibinfo{author}{\bibfnamefont{S.}~\bibnamefont{Hofling}}, \bibnamefont{and}
  \bibinfo{author}{\bibfnamefont{M.~C.} \bibnamefont{Gather}},
  \bibinfo{journal}{Science Advances} \textbf{\bibinfo{volume}{2}},
  \bibinfo{pages}{1600666} (\bibinfo{year}{2016}).

\bibitem[{\citenamefont{Liscidini et~al.}(2011)\citenamefont{Liscidini, Gerace,
  Sanvitto, and Bajoni}}]{Liscidini2011}
\bibinfo{author}{\bibfnamefont{M.}~\bibnamefont{Liscidini}},
  \bibinfo{author}{\bibfnamefont{D.}~\bibnamefont{Gerace}},
  \bibinfo{author}{\bibfnamefont{D.}~\bibnamefont{Sanvitto}}, \bibnamefont{and}
  \bibinfo{author}{\bibfnamefont{D.}~\bibnamefont{Bajoni}},
  \bibinfo{journal}{Appl. Phys. Lett.} \textbf{\bibinfo{volume}{98}},
  \bibinfo{pages}{121118} (\bibinfo{year}{2011}).

\bibitem[{\citenamefont{Pirotta et~al.}(2014)\citenamefont{Pirotta, Patrini,
  Liscidini, Galli, Dacarro, Canazza, Guizzeti, Comoretto, and
  Bajoni}}]{Pirotta2014}
\bibinfo{author}{\bibfnamefont{S.}~\bibnamefont{Pirotta}},
  \bibinfo{author}{\bibfnamefont{M.}~\bibnamefont{Patrini}},
  \bibinfo{author}{\bibfnamefont{M.}~\bibnamefont{Liscidini}},
  \bibinfo{author}{\bibfnamefont{M.}~\bibnamefont{Galli}},
  \bibinfo{author}{\bibfnamefont{G.}~\bibnamefont{Dacarro}},
  \bibinfo{author}{\bibfnamefont{G.}~\bibnamefont{Canazza}},
  \bibinfo{author}{\bibfnamefont{G.}~\bibnamefont{Guizzeti}},
  \bibinfo{author}{\bibfnamefont{D.}~\bibnamefont{Comoretto}},
  \bibnamefont{and} \bibinfo{author}{\bibfnamefont{D.}~\bibnamefont{Bajoni}},
  \bibinfo{journal}{Appl. Phys. Lett.} \textbf{\bibinfo{volume}{104}},
  \bibinfo{pages}{051111} (\bibinfo{year}{2014}).

\bibitem[{\citenamefont{Lerario et~al.}(2017)\citenamefont{Lerario, Ballarini,
  Fieramosca, Cannavale, Genco, Mangione, Gambino, Dominici, De~Giorgi, Gigli
  et~al.}}]{Lerario2017}
\bibinfo{author}{\bibfnamefont{G.}~\bibnamefont{Lerario}},
  \bibinfo{author}{\bibfnamefont{D.}~\bibnamefont{Ballarini}},
  \bibinfo{author}{\bibfnamefont{A.}~\bibnamefont{Fieramosca}},
  \bibinfo{author}{\bibfnamefont{A.}~\bibnamefont{Cannavale}},
  \bibinfo{author}{\bibfnamefont{A.}~\bibnamefont{Genco}},
  \bibinfo{author}{\bibfnamefont{F.}~\bibnamefont{Mangione}},
  \bibinfo{author}{\bibfnamefont{S.}~\bibnamefont{Gambino}},
  \bibinfo{author}{\bibfnamefont{L.}~\bibnamefont{Dominici}},
  \bibinfo{author}{\bibfnamefont{M.}~\bibnamefont{De~Giorgi}},
  \bibinfo{author}{\bibfnamefont{G.}~\bibnamefont{Gigli}},
  \bibnamefont{et~al.}, \bibinfo{journal}{Light: Science \& Applications}
  \textbf{\bibinfo{volume}{6}}, \bibinfo{pages}{16212} (\bibinfo{year}{2017}).

\bibitem[{\citenamefont{Solnyshkov et~al.}(2014)\citenamefont{Solnyshkov,
  Tercas, and Malpuech}}]{Solnyshkov2014}
\bibinfo{author}{\bibfnamefont{D.~D.} \bibnamefont{Solnyshkov}},
  \bibinfo{author}{\bibfnamefont{H.}~\bibnamefont{Tercas}}, \bibnamefont{and}
  \bibinfo{author}{\bibfnamefont{G.}~\bibnamefont{Malpuech}},
  \bibinfo{journal}{Appl. Phys. Lett.} \textbf{\bibinfo{volume}{105}},
  \bibinfo{pages}{231102} (\bibinfo{year}{2014}).

\bibitem[{\citenamefont{Walker et~al.}(2013)\citenamefont{Walker, Tinkler,
  Durska, Whittaker, Luxmoore, Royall, Krizhanovskii, Skolnick, Farrer, and
  Ritchie}}]{Walker2013}
\bibinfo{author}{\bibfnamefont{P.~M.} \bibnamefont{Walker}},
  \bibinfo{author}{\bibfnamefont{L.}~\bibnamefont{Tinkler}},
  \bibinfo{author}{\bibfnamefont{M.}~\bibnamefont{Durska}},
  \bibinfo{author}{\bibfnamefont{D.~M.} \bibnamefont{Whittaker}},
  \bibinfo{author}{\bibfnamefont{I.~J.} \bibnamefont{Luxmoore}},
  \bibinfo{author}{\bibfnamefont{B.}~\bibnamefont{Royall}},
  \bibinfo{author}{\bibfnamefont{D.~N.} \bibnamefont{Krizhanovskii}},
  \bibinfo{author}{\bibfnamefont{M.~S.} \bibnamefont{Skolnick}},
  \bibinfo{author}{\bibfnamefont{I.}~\bibnamefont{Farrer}}, \bibnamefont{and}
  \bibinfo{author}{\bibfnamefont{D.~A.} \bibnamefont{Ritchie}},
  \bibinfo{journal}{Appl. Phys. Lett.} \textbf{\bibinfo{volume}{102}},
  \bibinfo{pages}{012109} (\bibinfo{year}{2013}).

\bibitem[{\citenamefont{Rosenberg et~al.}(2016)\citenamefont{Rosenberg,
  Mazuz-Harpaz, Rapaport, West, and Pfeiffer}}]{Rosenberg2016}
\bibinfo{author}{\bibfnamefont{I.}~\bibnamefont{Rosenberg}},
  \bibinfo{author}{\bibfnamefont{Y.}~\bibnamefont{Mazuz-Harpaz}},
  \bibinfo{author}{\bibfnamefont{R.}~\bibnamefont{Rapaport}},
  \bibinfo{author}{\bibfnamefont{K.}~\bibnamefont{West}}, \bibnamefont{and}
  \bibinfo{author}{\bibfnamefont{L.}~\bibnamefont{Pfeiffer}},
  \bibinfo{journal}{Phys. Rev. B} \textbf{\bibinfo{volume}{93}},
  \bibinfo{pages}{195151} (\bibinfo{year}{2016}),
  \urlprefix\url{https://link.aps.org/doi/10.1103/PhysRevB.93.195151}.

\bibitem[{\citenamefont{Ellenbogen and Crozier}(2011)}]{Ellenboger2011}
\bibinfo{author}{\bibfnamefont{T.}~\bibnamefont{Ellenbogen}} \bibnamefont{and}
  \bibinfo{author}{\bibfnamefont{K.~B.} \bibnamefont{Crozier}},
  \bibinfo{journal}{Phys. Rev. B} \textbf{\bibinfo{volume}{84}},
  \bibinfo{pages}{161304} (\bibinfo{year}{2011}),
  \urlprefix\url{https://link.aps.org/doi/10.1103/PhysRevB.84.161304}.

\bibitem[{\citenamefont{Ciers et~al.}(2017)\citenamefont{Ciers, Roch, Carlin,
  Jacopin, Butt\'e, and Grandjean}}]{Ciers2017}
\bibinfo{author}{\bibfnamefont{J.}~\bibnamefont{Ciers}},
  \bibinfo{author}{\bibfnamefont{J.~G.} \bibnamefont{Roch}},
  \bibinfo{author}{\bibfnamefont{J.-F.} \bibnamefont{Carlin}},
  \bibinfo{author}{\bibfnamefont{G.}~\bibnamefont{Jacopin}},
  \bibinfo{author}{\bibfnamefont{R.}~\bibnamefont{Butt\'e}}, \bibnamefont{and}
  \bibinfo{author}{\bibfnamefont{N.}~\bibnamefont{Grandjean}},
  \bibinfo{journal}{Phys. Rev. Applied} \textbf{\bibinfo{volume}{7}},
  \bibinfo{pages}{034019} (\bibinfo{year}{2017}),
  \urlprefix\url{https://link.aps.org/doi/10.1103/PhysRevApplied.7.034019}.

\bibitem[{\citenamefont{Hu et~al.}(2017)\citenamefont{Hu, Luan, Scott, Yan,
  Mandrus, Xu, and Fei}}]{Hu2017}
\bibinfo{author}{\bibfnamefont{F.}~\bibnamefont{Hu}},
  \bibinfo{author}{\bibfnamefont{Y.}~\bibnamefont{Luan}},
  \bibinfo{author}{\bibfnamefont{M.~E.} \bibnamefont{Scott}},
  \bibinfo{author}{\bibfnamefont{J.}~\bibnamefont{Yan}},
  \bibinfo{author}{\bibfnamefont{D.~G.} \bibnamefont{Mandrus}},
  \bibinfo{author}{\bibfnamefont{X.}~\bibnamefont{Xu}}, \bibnamefont{and}
  \bibinfo{author}{\bibfnamefont{Z.}~\bibnamefont{Fei}},
  \bibinfo{journal}{Nature Photonics} \textbf{\bibinfo{volume}{11}},
  \bibinfo{pages}{356} (\bibinfo{year}{2017}).

\bibitem[{\citenamefont{Walker et~al.}(2015)\citenamefont{Walker, Tinkler,
  Skryabin, Yulin, Royall, Farrer, Ritchie, Skolnick, and
  Krizhanovskii}}]{Walker2015}
\bibinfo{author}{\bibfnamefont{P.~M.} \bibnamefont{Walker}},
  \bibinfo{author}{\bibfnamefont{L.}~\bibnamefont{Tinkler}},
  \bibinfo{author}{\bibfnamefont{D.~V.} \bibnamefont{Skryabin}},
  \bibinfo{author}{\bibfnamefont{A.}~\bibnamefont{Yulin}},
  \bibinfo{author}{\bibfnamefont{B.}~\bibnamefont{Royall}},
  \bibinfo{author}{\bibfnamefont{I.}~\bibnamefont{Farrer}},
  \bibinfo{author}{\bibfnamefont{D.~A.} \bibnamefont{Ritchie}},
  \bibinfo{author}{\bibfnamefont{M.~S.} \bibnamefont{Skolnick}},
  \bibnamefont{and} \bibinfo{author}{\bibfnamefont{D.~N.}
  \bibnamefont{Krizhanovskii}}, \bibinfo{journal}{Nature Communications}
  \textbf{\bibinfo{volume}{6}}, \bibinfo{pages}{8317} (\bibinfo{year}{2015}).

\bibitem[{sup()}]{suppl}
\bibinfo{note}{See Supplemental Material at [URL will be inserted by
  publisher].}

\bibitem[{\citenamefont{Zuniga-Perez et~al.}(2016)\citenamefont{Zuniga-Perez,
  Kappei, Deparis, Reveret, Grundmann, de~Prado, Jamadi, Leymarie, Chenot, and
  Leroux}}]{Perez2016}
\bibinfo{author}{\bibfnamefont{J.}~\bibnamefont{Zuniga-Perez}},
  \bibinfo{author}{\bibfnamefont{L.}~\bibnamefont{Kappei}},
  \bibinfo{author}{\bibfnamefont{C.}~\bibnamefont{Deparis}},
  \bibinfo{author}{\bibfnamefont{F.}~\bibnamefont{Reveret}},
  \bibinfo{author}{\bibfnamefont{M.}~\bibnamefont{Grundmann}},
  \bibinfo{author}{\bibfnamefont{E.}~\bibnamefont{de~Prado}},
  \bibinfo{author}{\bibfnamefont{O.}~\bibnamefont{Jamadi}},
  \bibinfo{author}{\bibfnamefont{J.}~\bibnamefont{Leymarie}},
  \bibinfo{author}{\bibfnamefont{S.}~\bibnamefont{Chenot}}, \bibnamefont{and}
  \bibinfo{author}{\bibfnamefont{M.}~\bibnamefont{Leroux}},
  \bibinfo{journal}{Appl. Phys. Lett.} \textbf{\bibinfo{volume}{108}},
  \bibinfo{pages}{251904} (\bibinfo{year}{2016}).

\bibitem[{\citenamefont{Huang et~al.}(2001)\citenamefont{Huang, Mao, Feick,
  Yan, Wu, Kind, Weber, Russo, and Yang}}]{Huang2001}
\bibinfo{author}{\bibfnamefont{M.~H.} \bibnamefont{Huang}},
  \bibinfo{author}{\bibfnamefont{S.}~\bibnamefont{Mao}},
  \bibinfo{author}{\bibfnamefont{H.}~\bibnamefont{Feick}},
  \bibinfo{author}{\bibfnamefont{H.}~\bibnamefont{Yan}},
  \bibinfo{author}{\bibfnamefont{Y.}~\bibnamefont{Wu}},
  \bibinfo{author}{\bibfnamefont{H.}~\bibnamefont{Kind}},
  \bibinfo{author}{\bibfnamefont{E.}~\bibnamefont{Weber}},
  \bibinfo{author}{\bibfnamefont{R.}~\bibnamefont{Russo}}, \bibnamefont{and}
  \bibinfo{author}{\bibfnamefont{P.}~\bibnamefont{Yang}},
  \bibinfo{journal}{Science} \textbf{\bibinfo{volume}{292}},
  \bibinfo{pages}{1897} (\bibinfo{year}{2001}).

\bibitem[{\citenamefont{Zamfirescu et~al.}(2002)\citenamefont{Zamfirescu,
  Kavokin, Gil, Malpuech, and Kaliteevski}}]{Zamfirescu2002}
\bibinfo{author}{\bibfnamefont{M.}~\bibnamefont{Zamfirescu}},
  \bibinfo{author}{\bibfnamefont{A.}~\bibnamefont{Kavokin}},
  \bibinfo{author}{\bibfnamefont{B.}~\bibnamefont{Gil}},
  \bibinfo{author}{\bibfnamefont{G.}~\bibnamefont{Malpuech}}, \bibnamefont{and}
  \bibinfo{author}{\bibfnamefont{M.}~\bibnamefont{Kaliteevski}},
  \bibinfo{journal}{Phys. Rev. B} \textbf{\bibinfo{volume}{65}},
  \bibinfo{pages}{161205} (\bibinfo{year}{2002}),
  \urlprefix\url{https://link.aps.org/doi/10.1103/PhysRevB.65.161205}.

\bibitem[{\citenamefont{Chu et~al.}(2008)\citenamefont{Chu, Olmedo, Yang, Kong,
  and Liu}}]{Chu2008}
\bibinfo{author}{\bibfnamefont{S.}~\bibnamefont{Chu}},
  \bibinfo{author}{\bibfnamefont{M.}~\bibnamefont{Olmedo}},
  \bibinfo{author}{\bibfnamefont{Z.}~\bibnamefont{Yang}},
  \bibinfo{author}{\bibfnamefont{J.}~\bibnamefont{Kong}}, \bibnamefont{and}
  \bibinfo{author}{\bibfnamefont{J.}~\bibnamefont{Liu}},
  \bibinfo{journal}{Appl. Phys. Lett.} \textbf{\bibinfo{volume}{93}},
  \bibinfo{pages}{181106} (\bibinfo{year}{2008}).

\bibitem[{\citenamefont{Vanmaekelbergh and van
  Vugt}(2011)}]{Vanmaekelbergh2011}
\bibinfo{author}{\bibfnamefont{D.}~\bibnamefont{Vanmaekelbergh}}
  \bibnamefont{and} \bibinfo{author}{\bibfnamefont{L.~K.} \bibnamefont{van
  Vugt}}, \bibinfo{journal}{Nanoscale} \textbf{\bibinfo{volume}{3}},
  \bibinfo{pages}{2783} (\bibinfo{year}{2011}).

\bibitem[{\citenamefont{Versteegh et~al.}(2011)\citenamefont{Versteegh, Kuis,
  Stoof, and Dijkhuis}}]{Versteegh2011}
\bibinfo{author}{\bibfnamefont{M.~A.~M.} \bibnamefont{Versteegh}},
  \bibinfo{author}{\bibfnamefont{T.}~\bibnamefont{Kuis}},
  \bibinfo{author}{\bibfnamefont{H.~T.~C.} \bibnamefont{Stoof}},
  \bibnamefont{and} \bibinfo{author}{\bibfnamefont{J.~I.}
  \bibnamefont{Dijkhuis}}, \bibinfo{journal}{Phys. Rev. B}
  \textbf{\bibinfo{volume}{84}}, \bibinfo{pages}{035207}
  (\bibinfo{year}{2011}),
  \urlprefix\url{https://link.aps.org/doi/10.1103/PhysRevB.84.035207}.

\bibitem[{\citenamefont{Versteegh et~al.}(2012)\citenamefont{Versteegh,
  Vanmaekelbergh, and Dijkhuis}}]{Versteegh2012}
\bibinfo{author}{\bibfnamefont{M.~A.~M.} \bibnamefont{Versteegh}},
  \bibinfo{author}{\bibfnamefont{D.}~\bibnamefont{Vanmaekelbergh}},
  \bibnamefont{and} \bibinfo{author}{\bibfnamefont{J.~I.}
  \bibnamefont{Dijkhuis}}, \bibinfo{journal}{Phys. Rev. Lett.}
  \textbf{\bibinfo{volume}{108}}, \bibinfo{pages}{157402}
  (\bibinfo{year}{2012}),
  \urlprefix\url{https://link.aps.org/doi/10.1103/PhysRevLett.108.157402}.

\bibitem[{\citenamefont{Levrat et~al.}(2010)\citenamefont{Levrat, Butt\'e,
  Feltin, Carlin, Grandjean, Solnyshkov, and Malpuech}}]{Levrat2010}
\bibinfo{author}{\bibfnamefont{J.}~\bibnamefont{Levrat}},
  \bibinfo{author}{\bibfnamefont{R.}~\bibnamefont{Butt\'e}},
  \bibinfo{author}{\bibfnamefont{E.}~\bibnamefont{Feltin}},
  \bibinfo{author}{\bibfnamefont{J.-F. m.~c.} \bibnamefont{Carlin}},
  \bibinfo{author}{\bibfnamefont{N.}~\bibnamefont{Grandjean}},
  \bibinfo{author}{\bibfnamefont{D.}~\bibnamefont{Solnyshkov}},
  \bibnamefont{and} \bibinfo{author}{\bibfnamefont{G.}~\bibnamefont{Malpuech}},
  \bibinfo{journal}{Phys. Rev. B} \textbf{\bibinfo{volume}{81}},
  \bibinfo{pages}{125305} (\bibinfo{year}{2010}),
  \urlprefix\url{https://link.aps.org/doi/10.1103/PhysRevB.81.125305}.

\bibitem[{\citenamefont{Jamadi et~al.}(2016)\citenamefont{Jamadi, R\'everet,
  Mallet, Disseix, M\'edard, Mihailovic, Solnyshkov, Malpuech, Leymarie,
  Lafosse et~al.}}]{Jamadi2016}
\bibinfo{author}{\bibfnamefont{O.}~\bibnamefont{Jamadi}},
  \bibinfo{author}{\bibfnamefont{F.}~\bibnamefont{R\'everet}},
  \bibinfo{author}{\bibfnamefont{E.}~\bibnamefont{Mallet}},
  \bibinfo{author}{\bibfnamefont{P.}~\bibnamefont{Disseix}},
  \bibinfo{author}{\bibfnamefont{F.}~\bibnamefont{M\'edard}},
  \bibinfo{author}{\bibfnamefont{M.}~\bibnamefont{Mihailovic}},
  \bibinfo{author}{\bibfnamefont{D.}~\bibnamefont{Solnyshkov}},
  \bibinfo{author}{\bibfnamefont{G.}~\bibnamefont{Malpuech}},
  \bibinfo{author}{\bibfnamefont{J.}~\bibnamefont{Leymarie}},
  \bibinfo{author}{\bibfnamefont{X.}~\bibnamefont{Lafosse}},
  \bibnamefont{et~al.}, \bibinfo{journal}{Phys. Rev. B}
  \textbf{\bibinfo{volume}{93}}, \bibinfo{pages}{115205}
  (\bibinfo{year}{2016}),
  \urlprefix\url{https://link.aps.org/doi/10.1103/PhysRevB.93.115205}.

\bibitem[{\citenamefont{Vi\~na et~al.}(1984)\citenamefont{Vi\~na, Logothetidis,
  and Cardona}}]{Vina1984}
\bibinfo{author}{\bibfnamefont{L.}~\bibnamefont{Vi\~na}},
  \bibinfo{author}{\bibfnamefont{S.}~\bibnamefont{Logothetidis}},
  \bibnamefont{and} \bibinfo{author}{\bibfnamefont{M.}~\bibnamefont{Cardona}},
  \bibinfo{journal}{Phys. Rev. B} \textbf{\bibinfo{volume}{30}},
  \bibinfo{pages}{1979} (\bibinfo{year}{1984}),
  \urlprefix\url{https://link.aps.org/doi/10.1103/PhysRevB.30.1979}.

\bibitem[{\citenamefont{Haug and Grob}(1967)}]{Haug1967}
\bibinfo{author}{\bibfnamefont{H.}~\bibnamefont{Haug}} \bibnamefont{and}
  \bibinfo{author}{\bibfnamefont{K.}~\bibnamefont{Grob}},
  \bibinfo{journal}{Phys. Lett. A} \textbf{\bibinfo{volume}{26}},
  \bibinfo{pages}{41} (\bibinfo{year}{1967}).

\bibitem[{\citenamefont{Schneider et~al.}(2013)\citenamefont{Schneider,
  Rahimi-Iman, Kim, Fischer, Savenko, Amthor, Lermer, Wolf, Worschech,
  Kulakovskii et~al.}}]{Schneider2013}
\bibinfo{author}{\bibfnamefont{C.}~\bibnamefont{Schneider}},
  \bibinfo{author}{\bibfnamefont{A.}~\bibnamefont{Rahimi-Iman}},
  \bibinfo{author}{\bibfnamefont{N.~Y.} \bibnamefont{Kim}},
  \bibinfo{author}{\bibfnamefont{J.}~\bibnamefont{Fischer}},
  \bibinfo{author}{\bibfnamefont{I.~G.} \bibnamefont{Savenko}},
  \bibinfo{author}{\bibfnamefont{M.}~\bibnamefont{Amthor}},
  \bibinfo{author}{\bibfnamefont{M.}~\bibnamefont{Lermer}},
  \bibinfo{author}{\bibfnamefont{A.}~\bibnamefont{Wolf}},
  \bibinfo{author}{\bibfnamefont{L.}~\bibnamefont{Worschech}},
  \bibinfo{author}{\bibfnamefont{V.~D.} \bibnamefont{Kulakovskii}},
  \bibnamefont{et~al.}, \bibinfo{journal}{Nature}
  \textbf{\bibinfo{volume}{497}}, \bibinfo{pages}{348} (\bibinfo{year}{2013}).

\bibitem[{\citenamefont{Kavokin and Malpuech}(2003)}]{CavityPol}
\bibinfo{author}{\bibfnamefont{A.}~\bibnamefont{Kavokin}} \bibnamefont{and}
  \bibinfo{author}{\bibfnamefont{G.}~\bibnamefont{Malpuech}},
  \emph{\bibinfo{title}{Cavity polaritons}} (\bibinfo{publisher}{Elsevier},
  \bibinfo{year}{2003}), ISBN \bibinfo{isbn}{978-0-12-533032-9}.

\bibitem[{\citenamefont{Tassone et~al.}(1997)\citenamefont{Tassone,
  Piermarocchi, Savona, Quattropani, and Schwendimann}}]{Tassone1997}
\bibinfo{author}{\bibfnamefont{F.}~\bibnamefont{Tassone}},
  \bibinfo{author}{\bibfnamefont{C.}~\bibnamefont{Piermarocchi}},
  \bibinfo{author}{\bibfnamefont{V.}~\bibnamefont{Savona}},
  \bibinfo{author}{\bibfnamefont{A.}~\bibnamefont{Quattropani}},
  \bibnamefont{and}
  \bibinfo{author}{\bibfnamefont{P.}~\bibnamefont{Schwendimann}},
  \bibinfo{journal}{Phys. Rev. B} \textbf{\bibinfo{volume}{56}},
  \bibinfo{pages}{7554} (\bibinfo{year}{1997}),
  \urlprefix\url{https://link.aps.org/doi/10.1103/PhysRevB.56.7554}.

\bibitem[{\citenamefont{Defrance et~al.}(2016)\citenamefont{Defrance, Lemaitre,
  Ajib, Benedicto, Mallet, Polles, Plumey, Mihailovic, Centeno, Ciraci
  et~al.}}]{Defrance2016}
\bibinfo{author}{\bibfnamefont{J.}~\bibnamefont{Defrance}},
  \bibinfo{author}{\bibfnamefont{C.}~\bibnamefont{Lemaitre}},
  \bibinfo{author}{\bibfnamefont{R.}~\bibnamefont{Ajib}},
  \bibinfo{author}{\bibfnamefont{J.}~\bibnamefont{Benedicto}},
  \bibinfo{author}{\bibfnamefont{E.}~\bibnamefont{Mallet}},
  \bibinfo{author}{\bibfnamefont{R.}~\bibnamefont{Polles}},
  \bibinfo{author}{\bibfnamefont{J.-P.} \bibnamefont{Plumey}},
  \bibinfo{author}{\bibfnamefont{M.}~\bibnamefont{Mihailovic}},
  \bibinfo{author}{\bibfnamefont{E.}~\bibnamefont{Centeno}},
  \bibinfo{author}{\bibfnamefont{C.}~\bibnamefont{Ciraci}},
  \bibnamefont{et~al.}, \bibinfo{journal}{Journal of Open Research Software}
  \textbf{\bibinfo{volume}{4}}, \bibinfo{pages}{13} (\bibinfo{year}{2016}).

\end{thebibliography}

\section{Supplemental Material}

The supplemental material is divided into four sections. Section I gives more details on the fabrication and the geometry of sample W2 (half-cavity). Section II shows the polariton dispersion below and at threshold, measured with high spatial selection which allows to remove the dispersiveless emission from the cracks. It also shows the raw data allowing to extract the blue shift of polariton modes versus pumping.
The emission properties of sample W1 at high pumping densities, and the corresponding numerical simulations based on the solution of semi-classical Boltzmann equations are discussed in Section III.  The method used to extract the dispersion for sample W2 and the power dependence of the emission are explained in Section IV.

\subsection{I Fabrication details and sample W2 geometry}
The sample W1 has been  grown by molecular beam epitaxy on m-plane bulk ZnO substrate. It  consists of a 1-µm thick Zn$_{1-x}$Mg$_x$O (x=0.26) buffer and lower cladding layer, a 50-nm thick ZnO layer, and a 100-nm thick  Zn$_{1-x}$Mg$_x$O (x=0.26) upper cladding layer (see Fig.~1 of the main text). Thermal mismatch induces cracks as discussed in the main text. 

The first-order grating couplers have been fabricated by electron beam lithography using a 80-nm thick, negative-tone Hydrogen silsesquioxane resist. Each grating spans over a $100\times 100$ $\mu$m$^2$ area, and the grating grooves are oriented perpendicularly to the c-axis, that is, parallel to the thermal cracks. The target fill factor is fixed to $\sim 50$ \%, and the periods $\Lambda$ have been calculated so that the central value of the in-plane wavevector (33 $\mu$m$^{-1}$) is out-coupled  perpendicular to the sample surface ($\Lambda$ =180-220 nm). 

The modal confinement factor of the 1D TE$_0$ guided mode is estimated to be of $\sim 46$ \% in the ZnO guiding core while the overlapping with the grating is estimated to be lower than 1 \%. This leads to relatively weak extraction loss, estimated to be lower than 4\% of the guided power over a propagation distance of 40 $\mu$m (from 2D FDTD simulations).

The sample W2 consists of a bottom DBR comprising $30\times$(ZnO/ZnMgO) $\lambda/4$ layers (further details on the structural, morphological and optical characterization can be found in \cite{Perez2016}), and completed by a 130-nm thick ZnO layer (on top). A cross-section SEM image and the sketch of the sample W2 are shown in Figs.\~S1(a,b). Both samples display cracks, with a mean distance between them of 25 $\mu$m. The surface roughness for both samples was of the order of 1 nm, as obtained from $5\times 5$ $\mu$m$^2$ regions, scanned by atomic force microscopy. The precise values of the layer thicknesses (in the structures) were determined by combining cross- section secondary electron microscopy images and X-ray reflectivity.

\begin{figure}[tbp]
 \begin{center}
 \includegraphics[scale=0.5]{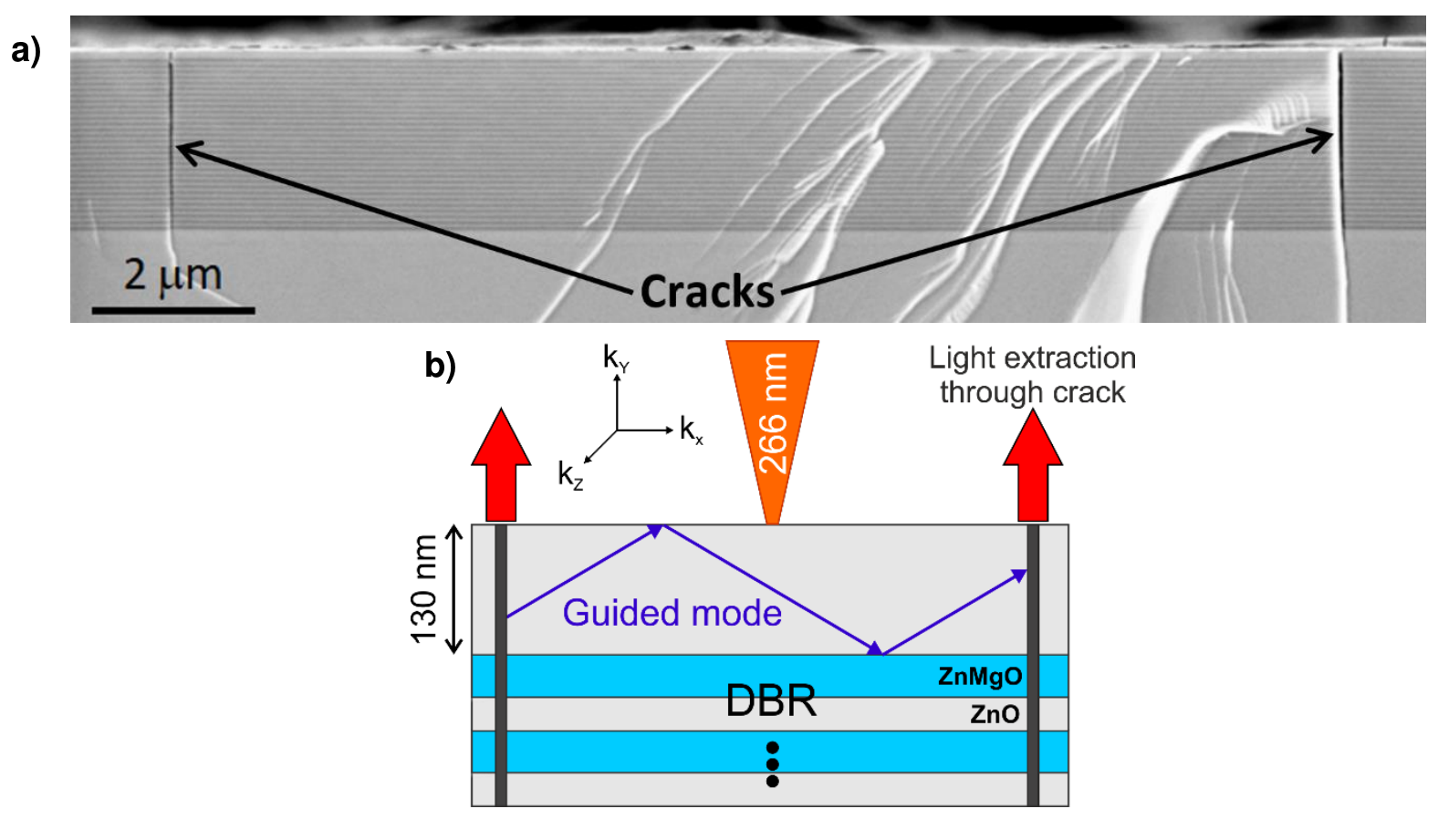}
 \caption{\label{figS1} Sample W2. a) SEM image of the half-cavity structure with clearly visible cracks. 
b) Sketch of a guided mode propagating in the active region. 
}
  \end{center}
 \end{figure}

\subsection{II Spatially selected emission and blue shift measurement}
The data shown in Fig.~2 of the main text are taken using a large pumping spot, covering the whole area of a horizontal cavity, which allows to minimize the lasing threshold. It makes difficult spatial selection of emission which therefore contains both light emitted from the grating area, which allows to observe the polariton dispersion, and also light from the cracks, which  appears as non-dispersive "horizontal lines" due to diffraction. Figure S2 shows measurements taken at 300 K, both below and at threshold, where a spatial selection of the grating emission by a pinhole is performed. The figure clearly shows the polariton dispersion whereas the dispersion-less emission from the cracks has disappeared.

\begin{figure}[tbp]
 \begin{center}
 \includegraphics[scale=0.4]{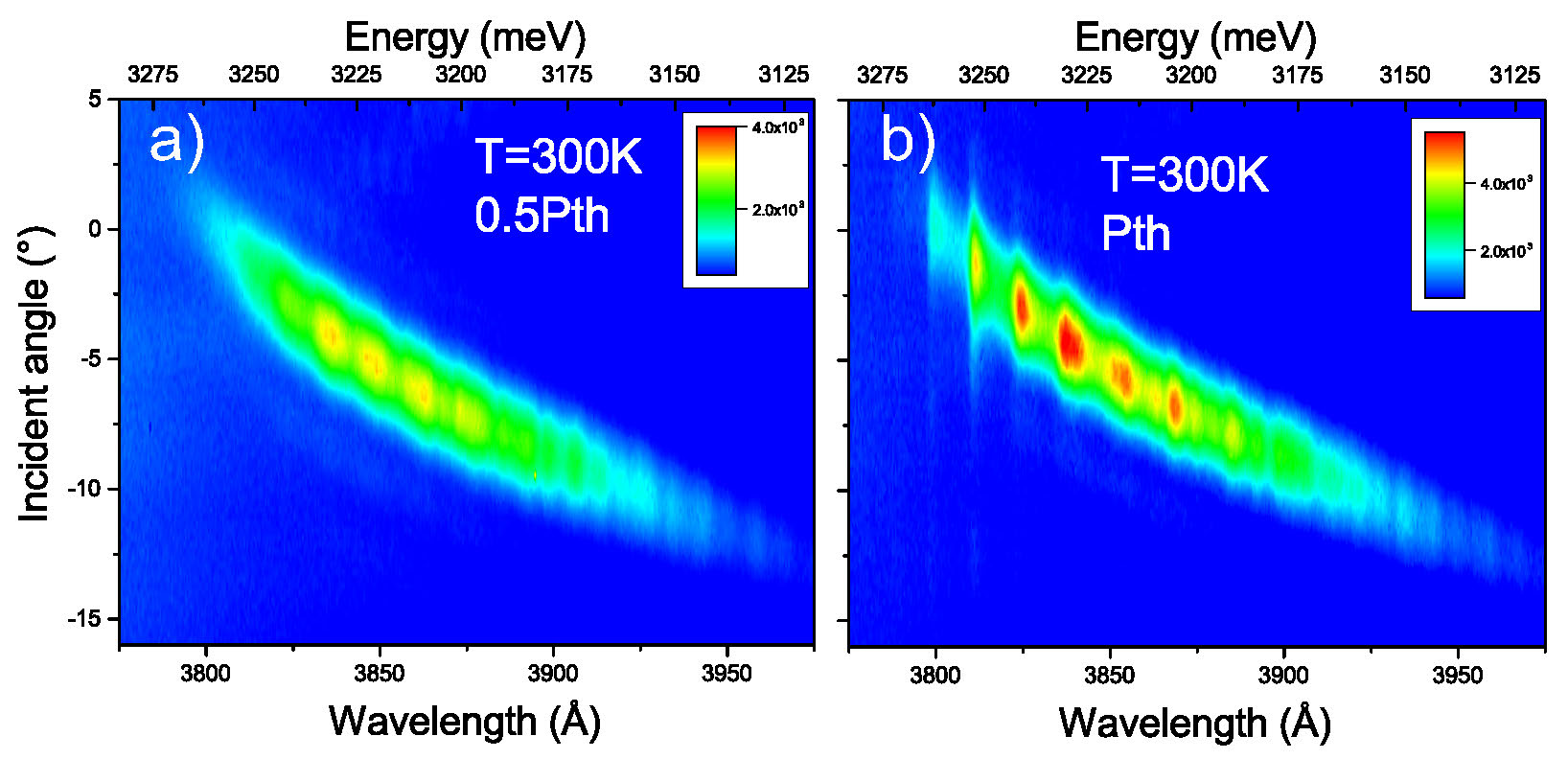}
 \caption{\label{figS2} Emission of W1 at 300 K versus energy and emission angle. The emission from the grating is selected by a pinhole which allows to remove the crack emission a) 0.5 $P_{th}$ b) $P_{th}$
}
  \end{center}
 \end{figure}

Figure S3 shows the PL spectra of the sample W1 without angular selection, versus pumping power at 5K and 300 K. These figures allows to follow the energy of the Fabry-Perot modes of the horizontal cavity versus pumping, and to track their blue shifts which are shown in Fig.~2(a,b) of the main text. 

\begin{figure}[tbp]
 \begin{center}
 \includegraphics[scale=0.33]{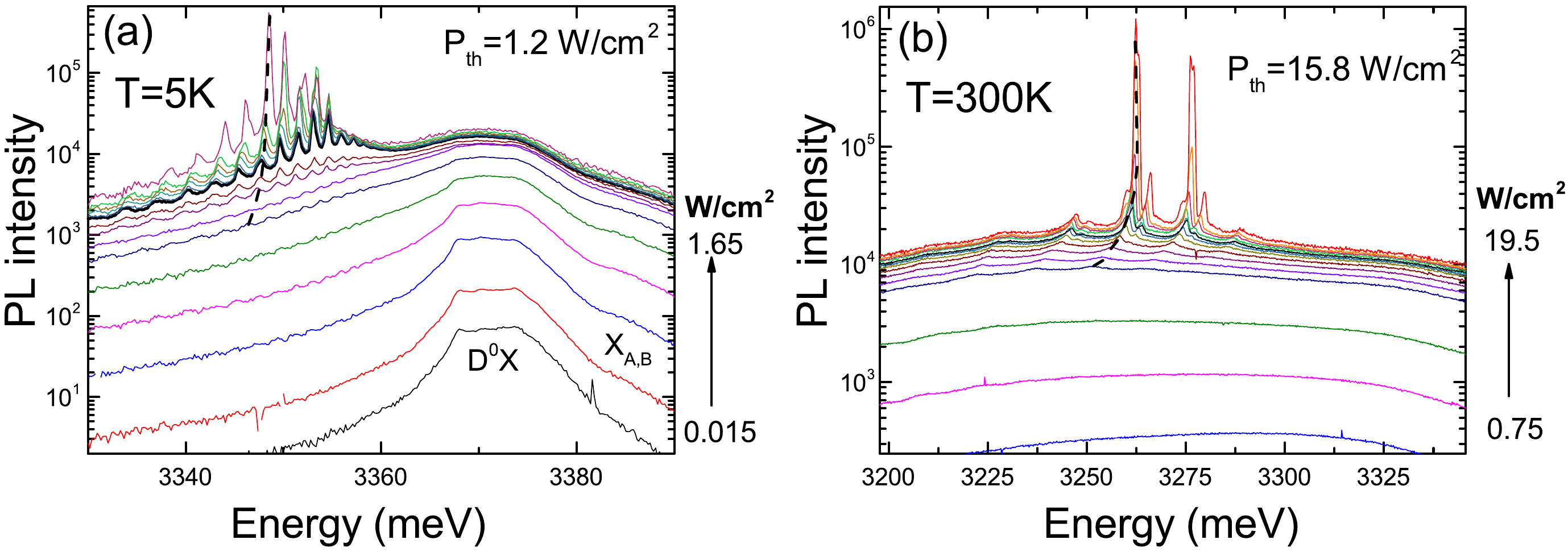}
 \caption{\label{figS3} Emission of W1 versus pumping power with spatial selection (grating only): a) 5 K b) 300 K
}
  \end{center}
 \end{figure}

\subsection{III High pumping densities}
The \textsf{movie.mp4} (also available at \url{https://www.youtube.com/watch?v=A0ZSpzKErZw}) shows the emission of W1, measured in the same conditions as those of Fig.~2(c,e) of the main text for a pumping power varying from 0.1 to 20 $P_{th}$. Two symmetric dispersion curves can be observed, corresponding  to  the  positive  and  negative directions of propagation. It is clear from this movie that the emission always fits the polariton dispersion (black curve), even if it becomes harder to determine at the largest pumping densities because of the strong mixing with the crack emission. The dispersion itself remains largely unperturbed, showing a slight blue shift of the order of 1-2 meV. The  lasing emission above threshold shows a clear red-shift, and corresponds to states with larger and larger photon fractions with increasing pumping power. The figure S4 is obtained from the same data as the movie, showing the emission spectrum versus energy taken at different pumping. 

\begin{figure}[tbp]
 \begin{center}
 \includegraphics[scale=0.3]{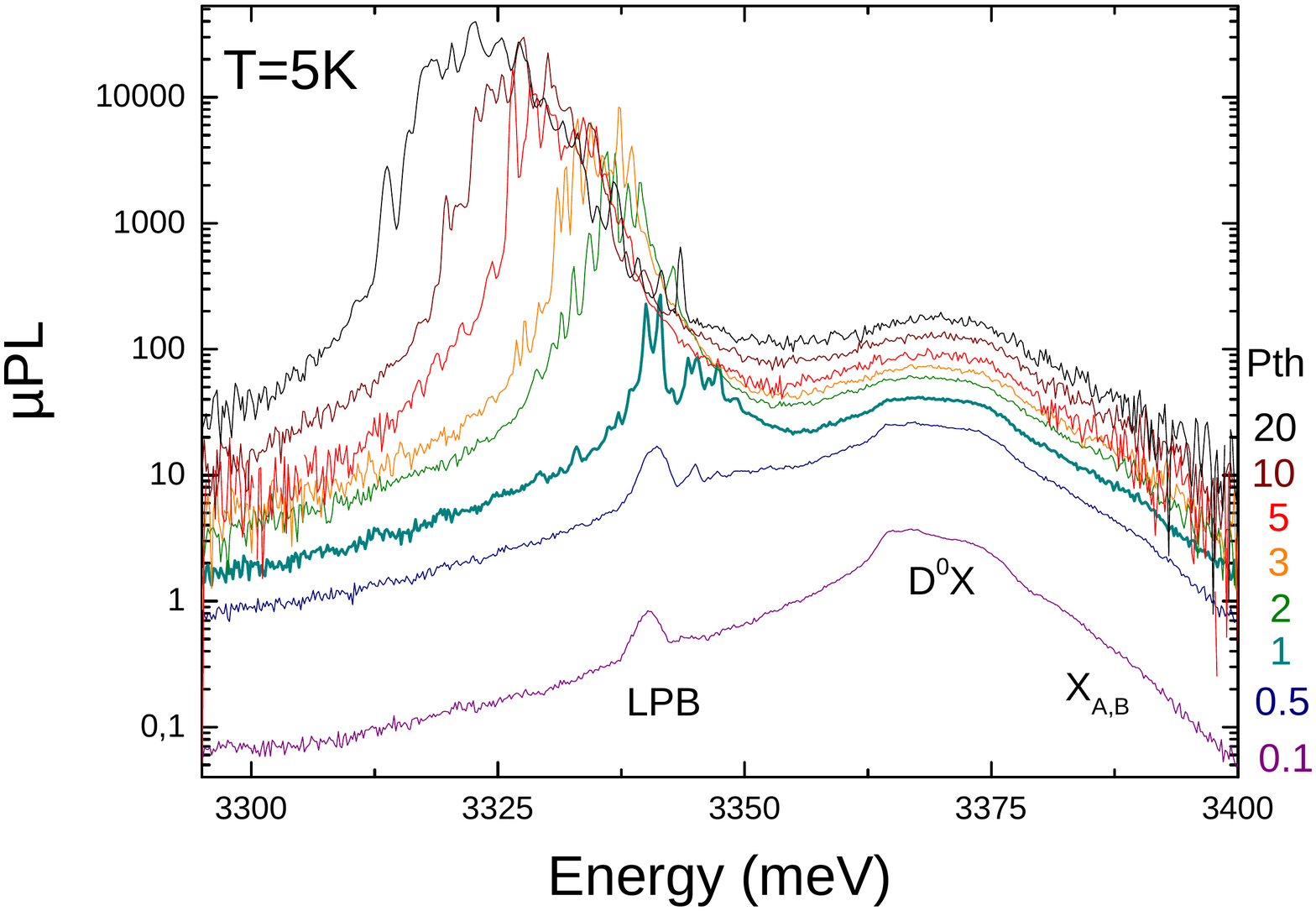}
 \caption{\label{figS4} PL of W1 at 5 K versus pumping. The data are the same as the one used for the movie. 
}
  \end{center}
 \end{figure}
 
 \begin{figure}[tbp]
 \begin{center}
 \includegraphics[scale=0.6]{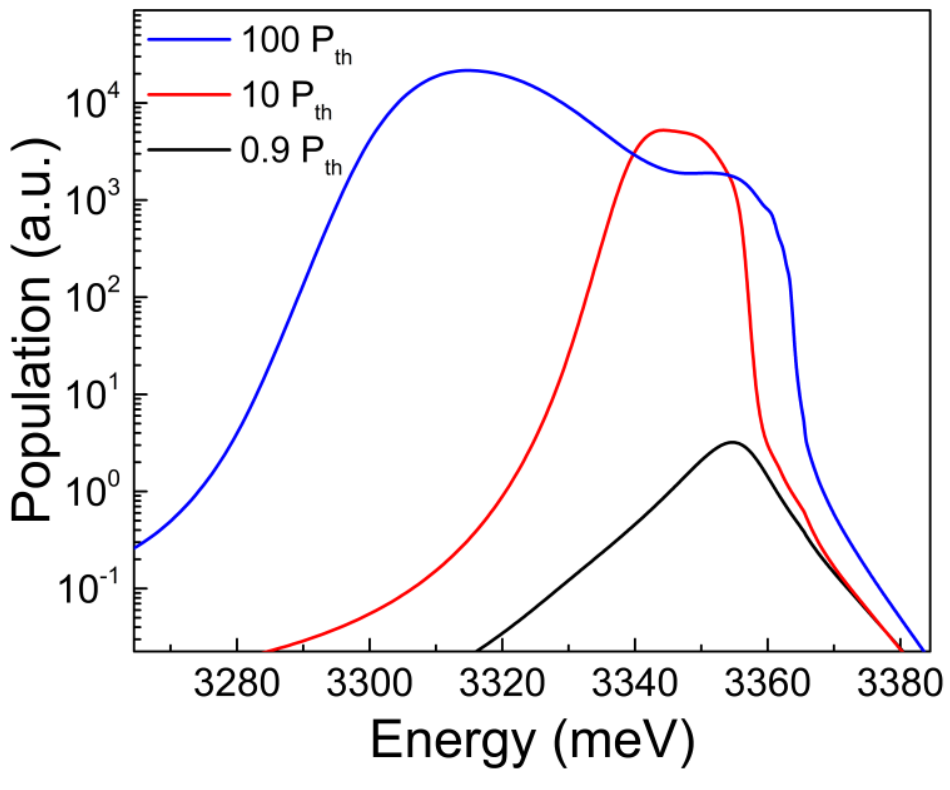}
 \caption{\label{figS5} Polariton distribution function versus energy, calculated as a numerical solution of the semi-classical Boltzmann equation for different pumping powers.}
\end{center}
\end{figure}

It compares qualitatively well with the Figure~S5 showing the results of simulations based on the numerical solution of the semi-classical Boltzmann equations for polaritons similar to the ones used in \cite{Solnyshkov2014}. The general formalism of semi-classical Boltzmann equations and its application to polaritons are discussed in \cite{CavityPol}.

\begin{eqnarray}
\frac{{d{n_{\bf{k}}}}}{{dt}} &=& {P_{\bf{k}}} - {\Gamma _{\bf{k}}}{n_{\bf{k}}} - {n_{\bf{k}}}\sum\limits_{{\bf{k}}'} {{W_{{\bf{k}} \to {\bf{k}}'}}\left( {{n_{{\bf{k}}'}} + 1} \right)}\\  &+& \left( {{n_{\bf{k}}} + 1} \right)\sum\limits_{{\bf{k}}'} {{W_{{\bf{k}}' \to {\bf{k}}}}{n_{{\bf{k}}'}}}\nonumber
\end{eqnarray}

Here, $n_\mathbf{k}$   is the population of polariton or exciton states at a given wavevector, $P_\mathbf{k}$  is the pumping (we represent the non-resonant pumping by injection of excitons, assuming that they rapidly thermalize with respect to the free exciton energy with optical phonons),  $\Gamma_\mathbf{k}$ are the decay rates, determined by the group velocity and the reflection on the cavity edges (we assume 50\% reflection probability, as confirmed by COMSOL simulations). 

$W$  are the scattering rates between the states, assisted by the exciton-phonon and exciton-exciton interactions. The exciton-phonon scattering rate can be written as:
\[\begin{gathered}
  W_{\vec k \to \vec k'}^{phon} = \frac{{2\pi }}{\hbar }\sum\limits_{\vec q} {{{\left| {M\left( {\vec q} \right)} \right|}^2}\left( {0,1 + N_q^{phon}} \right)}  \\ 
   \times \frac{{\hbar {\gamma _{k'}}/\pi }}{{{{\left( {E\left( {k'} \right) - E\left( k \right) \pm \hbar {\omega _q}} \right)}^2} + {{\left( {\hbar {\gamma _{k'}}} \right)}^2}}} \\ 
\end{gathered} \]
where $M(q)$ is the matrix element of interaction, depending on the phonon type (acoustic or optical), $0,1$ stand for phonon absorption or emission respectively, $N_{\vec{q}}^{phon}=1/(\exp(-E(\vec{q})/k_bT)-1)$ is the number of phonons with the energy given by the exchanged wavevector $q$,
 $\hbar\gamma_{k'}$ is the  broadening of the polariton states (induced by the lifetime or other sources). Thus, at higher temperatures the phonon-assisted processes become enhanced by the increase of the corresponding terms in this scattering rate. On the other hand, the exciton-exciton scattering rate strongly depends on the density of excitons:
\[\begin{gathered}
  W_{\vec k \to \vec k'}^{exc} = \frac{{2\pi }}{\hbar }\sum\limits_{\vec q} {{{\left| {{M_{ex}}} \right|}^2}N_{\vec q}^{exc}\left( {1 + N_{\vec q + \vec k' - \vec k}^{exc}} \right)}  \hfill \\
   \times \frac{{\hbar {\gamma _{k'}}/\pi }}{{{{\left( {E\left( {k'} \right) - E\left( k \right) + E\left( {q + k' - k} \right) - E\left( q \right)} \right)}^2} + {{\left( {\hbar {\gamma _{k'}}} \right)}^2}}} \hfill \\ 
\end{gathered} \]
We stress that these scattering rates concern only the excitonic fraction of the quasiparticles at any wavevector. Here, $M_{ex}$ is the matrix element of the exciton-exciton scattering, $N_{\vec{q}}^{exc}$ is the number of excitons (or exciton-polaritons) with a given wavevector. We see that the overall dependence of the scattering rate on the exciton density is quadratic, which greatly enhances the relaxation at high densities.

The Rabi splitting value is taken as 200 meV. The increase of pumping power enhances all scattering rates for two reasons: 1) the efficiency of exciton-exciton scattering increases with the density of excitons, and 2) bosonic stimulation increases all scattering rates. In planar vertical cavities this enhancement of scattering rate allows to overcome the well-known "bottleneck effect" \cite{Tassone1997}. In such cavities, the dispersion shows an energy minimum whose depth can be tuned by changing the exciton-photon detuning at zero in-plane wavevector. If the relaxation times are longer than the polariton lifetime near the ground state, polaritons cannot dissipate energy efficiently enough to reach the ground state during their lifetime and they can accumulate higher in energy in the bottleneck region. Increasing pumping power allows to enhance relaxation rates, and to overcome the bottleneck. Eventually, relaxation processes can become efficient enough to lead to the formation of a quasi-thermal distribution function and of a polariton Bose condensate in the dispersion ground state. However, one can engineer the dispersion by changing the exciton-photon detuning. By going to negative detuning, the energy dip in reciprocal space becomes deeper and sharper. It takes longer for polaritons to reach the ground state, and the more photon like polaritons have typically a shorter radiative lifetime. So the pumping density which was good enough to allow condensation in the ground state at positive exciton-photon detuning leads to the formation of a bottleneck at more negative detuning and a further increase of pumping will be needed for polaritons to relax down to the dispersion ground state at this detuning. This picture has been widely discussed theoretically and evidenced experimentally in a large amount of works \cite{Kasprzak2008,Levrat2010,Feng2013} and is well established. In a waveguide geometry, the picture is quite similar, except that there is no ground state. In this case, the competition between relaxation towards several modes defines the position of the "bottleneck", namely the states which are the most favorable from a loss-gain point of view. When the pumping power is increased, the enhancement of the scattering rates displaces the bottleneck region deeper in energy, leading to a continuous red-shift of the maximum of emission, as observed both experimentally and theoretically.

\subsection{IV Extraction of dispersion (sample W2), power dependence of emission}

The Fabry-Perot modes visible in Fig.~3 of the main text result from the horizontal confinement of light between the two cracks surrounding the excitation spot. These two cracks form an horizontal cavity. The energy spacing between the peaks is determined by the length of the cavity and the slope of the dispersion relation. By knowing the length of the cavity $L$, we can calculate the difference $\delta k_z=\pi/L$ between subsequent modes. Knowing the energy and the wavevector spacing between modes, one simply needs a reference wavevector $k_z$ to extract the experimental dispersion. It is approximately estimated by adjusting the experimental points to the theoretical dispersion. The result of this procedure is shown in Fig.~S6(a).
The theoretical dispersion of the guided modes of both samples was calculated using the scattering matrix formalism \cite{Defrance2016}  taking into account all the sample layers. We choose the bare photonic mode TE$_0$ by default. The refractive index of the ZnMgO alloy was measured by ellipsometry. The validity of the thickness and the refractive index were confirmed by reflectivity experiments and transfer matrix calculation. The exciton oscillator strength for ZnO has been chosen in agreement with \cite{Jamadi2016}.

\begin{figure}[tbp]
 \begin{center}
 \includegraphics[scale=0.3]{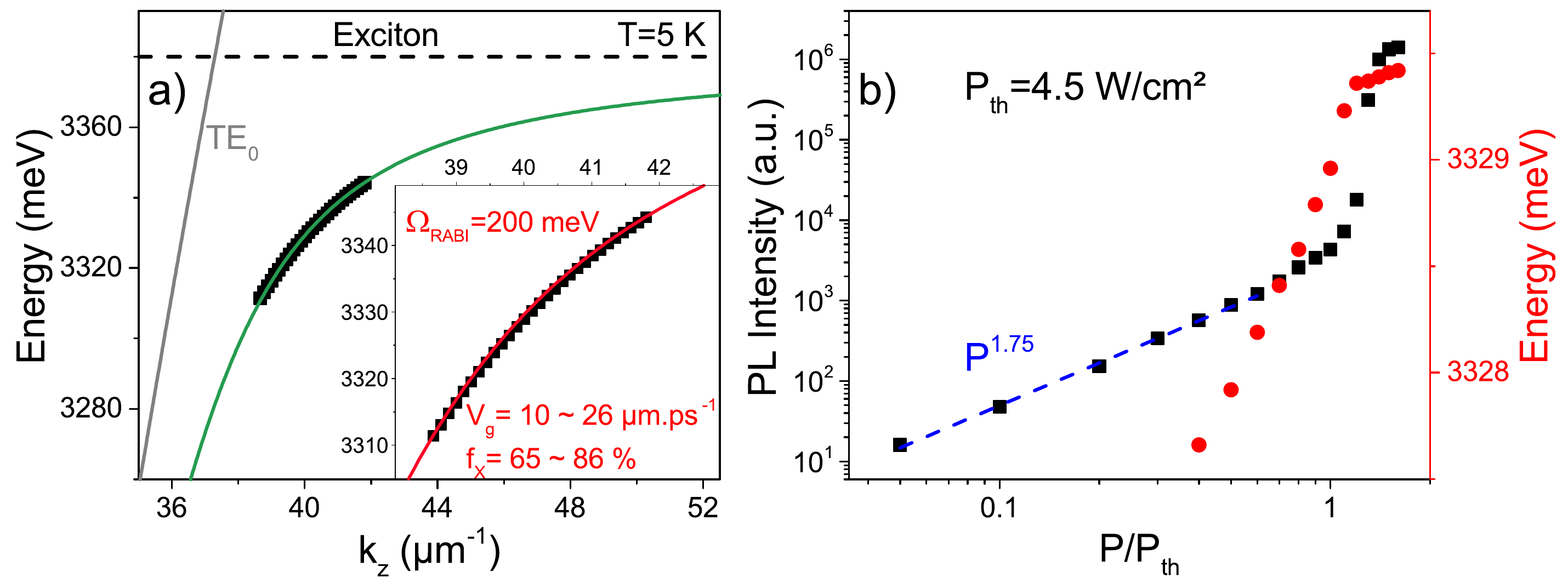}
 \caption{\label{figS6} Sample W2. Dispersion of the guided polariton mode and power dependence of emission. 
a) Experimental dispersion at 5 K extracted from interference patterns (black squares) and the theoretical dispersion calculated using scattering matrix formalism (green). The inset shows the same experimental dispersion fitted with a coupled oscillator model which allows to extract the group velocity and the excitonic fraction of the polariton.
b) Emission intensity (black squares) and energy (red dots) of the most intense lasing peak versus pumping power. The exponent of the slope fitting the emission intensity below the pumping threshold ($P_{th}$) lies between 1 and 2, which indicates a relaxation, assisted both by exciton-exciton and exciton-LO phonon interactions. }
  \end{center}
 \end{figure}
 
Another way to calculate the dispersion curve is based on the coupled oscillator model. It also provides a good agreement with experimental results and allows to determine the exciton and photon fractions of the amplified modes. The corresponding curves are shown in Fig.~2 of the main text for the sample W1 and in Fig.~S6(a) for the sample W2 (red curve on the inset).

The power dependence of emission extracted from the data presented in Fig.~3(a) of the main text is shown in Fig.~S6(b), together with the energy shift of the Fabry-Perot modes which is of the order of 1.5 meV at 2 $P_{th}$ (less than 1\% of the Rabi splitting $\Omega$).

\end{document}